\documentclass[preprint,preprintnumbers,amsmath,amssymb]{revtex4}
\usepackage{graphicx}
\usepackage{dcolumn}
\usepackage{bm}
\usepackage{rotating}
\begin{document}
\newcommand{\n}{\noindent}
\newcommand{\e}{\mbox{e}}

\title{On certain new integrable second order nonlinear differential equations and
their connection with two dimensional Lotka - Volterra system}
\author{R~Gladwin~Pradeep, V~K~Chandrasekar, M~Senthilvelan and M~Lakshmanan}

\affiliation {Centre for Nonlinear Dynamics, School of Physics,
Bharathidasan University, Tiruchirapalli - 620 024, India.}

\date{\today}
\begin{abstract}
In this paper, we consider a second order nonlinear ordinary differential equation of the form
$\ddot{x}+k_1\frac{\dot{x}^2}{x}+(k_2+k_3x)\dot{x}+k_4x^3+k_5x^2+k_6x=0$, where $k_i$'s, $i=1,2,...,6,$ are arbitrary
parameters.  By using the modified Prelle-Singer procedure, we identify five new integrable cases in this equation besides two
known integrable cases, namely (i) $k_2=0,\,k_3=0$ and (ii) $k_1=0,\,k_2=0,\,k_5=0$.  Among these five, four equations admit time dependent first
integrals and the remaining one admits time independent first integral.  From the time independent
first integral, nonstandard Hamiltonian structure is deduced thereby proving the Liouville sense of integrability.  In the case of time dependent integrals, we either
explicitly integrate the system or transform to a time-independent case and deduce
the underlying Hamiltonian structure.  We
also demonstrate that the above second order ordinary differential equation is intimately related to the two-dimensional Lotka-Volterra (LV) system. From the integrable
parameters of above nonlinear equation and all the known integrable cases of the latter can be deduced thereby.
\end{abstract}

\keywords{}
\maketitle


\section{Introduction and statement of results}
\label{sec1}
\subsection{Introduction}

In this paper we investigate the integrability properties associated with the second order
nonlinear ordinary differential equation (ODE) of the form
\begin{eqnarray}
\ddot{x}+k_1\frac{\dot{x}^2}{x}+(k_2+k_3x)\dot{x}+k_4x^3+k_5x^2+k_6x=0,
\label{eq6}
\end{eqnarray}
where $k_i$'s, $i=1,2,...,6,$ are arbitrary parameters and the over dot denotes
differentiation with respect to `$t$'.  Equation (\ref{eq6}) is a combination of two
different classes of equations, namely Li\'enard type equation ($k_1=0$) and equation with
quadratic friction ($k_2=k_3=0$).  Using the transformation $z=\dot{x}$ one can reduce
Eq. (\ref{eq6}) to the Abel equation of the second kind{\footnotesize$^{1,2}$}.  However, the
resultant equation is not integrable in general.  A set of integrable parametric choices of this
equation has been listed in Refs. 1 \& 2.  The importance of the study of Eq. (\ref{eq6})
arises from the fact that this equation contains many physically interesting equations such as the
modified Emden equation, unforced Duffing oscillator, Helmholtz oscillator, etc and is intimately
related to the much analyzed biological model, namely the two dimensional
Lotka-Volterra system{\footnotesize$^{3,4}$} (LV),
\begin{subequations}
\begin{eqnarray}
&&\dot {x}=x(a_{1}+b_{11}x+b_{12}y),\label {eq1a}\\
&&\dot {y}=y(a_{2}+b_{21}x+b_{22}y).\label {eq1b}
\end{eqnarray}
\label {eq1}
\end{subequations}
In order to study the integrable properties of Eq. (\ref{eq6}) we employ the modified Prelle-Singer procedure (PS) which has been intensively used to identify new integrable cases for several equations{\footnotesize$^{5,6}$}.  Using this procedure we identify several new integrable
parametric choices and deduce their corresponding integrals of motion.  In order to prove the integrability we either associate a conservative Hamiltonian structure to the equation constructed from the time independent integrals of motion, thereby proving Liouville sense of integrability, or explicitly integrate the integrals of motion to obtain general solution, proving complete integrability.

  During the past three decades
or so several attempts have been made to explore the integrable cases in the LV system
because of the immense importance of the problem in mathematical ecology and biology{\footnotesize$^{7-12}$}.  Integrals of motion of the LV system have been constructed for several parametric choices{\footnotesize$^{9,10,12-19}$}.  However, we find that all these parametric choices effectively reduce to one of the three general parametric choices for which integrals of motion have been reported{\footnotesize$^{9}$}.  Interestingly,
we recover all these three integrable cases  in addition to many subcases by comparing and expressing the parameters
appearing in (\ref{eq6}) in terms of the LV system parameters.  The results exactly coincide with the
reported ones in the literature.  Thus, as a by-product, we recover all the known integrable cases
of LV by investigating the integrability properties of Eq. (\ref{eq6}).  At this point
we stress the fact that as far as integrability is concerned one can extract complete information
about (\ref{eq1}) from (\ref{eq6}) and not vice-versa.  In other words when one tries to deduce
the integrable cases of (\ref{eq6}) from (\ref{eq1}) only partial results can be extracted.  That is one will loose a major portion of the results when we proceed in the reverse way (see
Sec. VI for more details).

\subsection{Results}
Our main results can be summarized as follows.
Using the modified PS method we identify seven integrable parametric choices of which five seem to be new as far as our knowledge goes.  The underlying forms of the equations are
\begin{eqnarray}
&&\hspace{-1.2cm}\ddot{x}+k_1
\frac{\dot{x}^2}{x}+(k_2+ k_3x)\dot{x} +
k_4x^3+\frac{k_2k_4(3+2k_1)}{k_3(1+k_1)}x^2
+\frac{k_4k_2^2(2+k_1)}{k_3^2(1+k_1)}x=0,\label{tid5}\\
&&\hspace{-1.2cm}\ddot{x}+k_1\frac{\dot{x}^2}{x}+(k_2+k_3x)\dot{x}
+\frac{(1+k_1)k_3^2}{(3+2k_1)^2}x^3+\frac{k_2k_3}{(3+2k_1)}x^2+k_6x=0,\label{caseia,b}\\
&&\hspace{-1.2cm}\ddot{x}+k_1 \frac{\dot{x}^2}{x}+\left(k_2+k_3x\right)\dot{x}+\frac{k_3k_2(1+k_1)}{(3+2k_1)} x^2
+\frac{(2+k_1)k_2^2}{(3+2k_1)^2} x=0,\label{td2-1}\\
&&\hspace{-1.2cm}\ddot{x}+k_1 \frac{\dot{x}^2}{x}+k_2\dot{x}+k_5 x^2+\frac{2(3+2 k_1)k_2^2}{(5+4k_1)^2} x=0,\label{td3eq}\\
&&\hspace{-1.2cm}\ddot{x}+k_1\frac{\dot{x}^2}{x}+(k_2+k_3x)\dot{x}+k_4x^3+\frac{k_2k_3}{(3+2k_1)} x^2
+\frac{(2+k_1)k_2^2}{(3+2k_1)^2}x=0.\label{eqiv}
\end{eqnarray}
For these five cases we prove integrability either in the Liouville sense by constructing
the time independent Hamiltonian structure or deduce the
solution by explicitly integrating the time dependent integrals of motion.

The remaining two cases which are already known in the literature are
\begin{eqnarray}
&&\ddot{x}+k_1 \frac{\dot{x}^2}{x}+k_4 x^3+k_5 x^2+k_6 x=0\,\label{bern1}\\
&&\hspace{-5cm}\mbox{and}\nonumber\\
&&\ddot{x}+k_3 x\dot{x}+k_4 x^3+k_6x=0\,\label{tid4}.
\end{eqnarray}
We note that equation (\ref{bern1}) can be reduced to the Bernoulli's equation, which can then be integrated
to give the solution in terms of quadratures.  Equation (\ref{tid4}) is the modified Emden type equation with linear forcing term which has been studied in some detail in Refs. 20 \& 21.  The Hamiltonian
structure for this equation for the parametric choices $k_3^2\ge8k_4$ has been given in Ref. 20.
Here we present the Hamiltonian structure for the parametric choice $k_3^2<8k_4$ as well.  In addition to these seven integrable parameteric choices we find the following equation
\begin{eqnarray}
\ddot{x}+k_1 \frac{\dot{x}^2}{x}+(k_2+k_3x)
\dot{x}+\frac{k_3(k_2\pm\omega)}{2(2+k_1)} x^2+k_6
x=0,\label{eq87}
\end{eqnarray}
where $\omega=\sqrt{k_2^2-4(1+k_1)k_6}$, for which we obtain a
time-dependent integral of motion.
However, we are able to prove its complete integrability only with an additional parametric restriction $k_5=\frac{k_3k_2(1+k_1)}{(3+2k_1)}$, and $k_6=\frac{(2+k_1)k_2^2}{(3+2k_1)^2}$, which reduces Eq. (\ref{eq87}) to Eq. (\ref{td2-1}).  In this parametric choice we are able to explicitly integrate the time-dependent integral of motion to find the general solution.

The plan of the paper is as follows.  In the following section we briefly describe the
extended Prelle-Singer procedure{\footnotesize$^{5,6}$} applicable to second-order ODEs.  In Sec. \ref{sectid}
we identify the integrable parametric choices of Eq. (\ref{eq6})  which admit time independent integral of motion through the extended
PS procedure.  In Sec. IV we construct explicit conservative Hamiltonians from the time independent integral of motion.  Further we transform these Hamiltonians to simpler forms using canonical transformations in order to explicitly integrate the canonical equations of motion. In Sec. \ref{sectd}, we then identify the integrable cases of (\ref{eq6}) which admit explicit time-dependent
integrals of motion.  To establish the complete integrability of these cases
we transform the time-dependent integrals of motion
into time-independent integrals of motion and integrate the latter and derive the
general solution.  In Sec. \ref{connection} we present the connection between Eq. (\ref{eq6}) and the LV equation.  In Sec. VII we rewrite the results in terms of the LV parameters
and point out the integrable equations.  Finally, we present our conclusions in Sec. \ref{seccon}.

\section{Extended Prelle-Singer (PS) procedure}
\label{sec2}
\n Let us rewrite Eq.~(\ref{eq6}) in the form
\begin{eqnarray}
\ddot{x}=-(k_1\frac{\dot{x}^2}{x}+(k_2+k_3x)\dot{x}+k_4x^{3}+k_5x^{2}+k_6x)
\equiv \phi(x,\dot{x}).
 \label{met1}
\end{eqnarray}
Further, we assume that Eq. (\ref{met1})
admits a first integral $I(t,x,\dot{x})=C,$ with $C$ constant on the
solutions, so that the total differential becomes
\begin{eqnarray}
dI={I_t}{dt}+{I_{x}}{dx}+{I_{\dot{x}}{d\dot{x}}}=0,
\label{met3}
\end{eqnarray}
where subscript denotes partial differentiation with respect
to that variable. Rewriting Eq.~(\ref{met1}) in the form
$\phi dt-d\dot{x}=0$ and adding a null term
$S(t,x,\dot{x})\dot{x}dt - S(t,x,\dot{x})dx$ to the latter, we obtain that on
the solutions the 1-form
\begin{eqnarray}
(\phi +\dot{x}S) dt-Sdx-d\dot{x} = 0.
\label{met6}
\end{eqnarray}
Hence, on the solutions, the 1-forms (\ref{met3}) and
(\ref{met6}) must be proportional.  Multiplying (\ref{met6}) by the
function $ R(t,x,\dot{x})$ which acts as the integrating factor
for (\ref{met6}), we have on the solutions that
\begin{eqnarray}
dI=R(\phi+S\dot{x})dt-RSdx-Rd\dot{x}=0.
\label{met7}
\end{eqnarray}
Comparing Eq.~(\ref{met3})
with (\ref{met7}), we have the relations
\begin{eqnarray}
 I_{t}  = R(\phi+\dot{x}S),\quad
 I_{x}  = -RS, \quad
 I_{\dot{x}}  = -R.
 \label{met8}
\end{eqnarray}
Then the compatibility conditions,
$I_{tx}=I_{xt}$, $I_{t\dot{x}}=I_{{\dot{x}}t}$, $I_{x{\dot{x}}}=I_{{\dot{x}}x}$,
between the different equations of (\ref{met8}), provide us the relations
\begin{eqnarray}
S_t+\dot{x}S_x+\phi S_{\dot{x}} &=&
   -\phi_x+\phi_{\dot{x}}S+S^2,\label {lin02}\\
R_t+\dot{x}R_x+\phi R_{\dot{x}} & =&
-(\phi_{\dot{x}}+S)R,\label {lin03}\\
R_x-SR_{\dot{x}}-RS_{\dot{x}}  &= &0.
\qquad \qquad\qquad \;\;\;\label {lin04}
\end{eqnarray}

Solving Eqs.~(\ref{lin02})-(\ref{lin04}) one can obtain expressions for $S$ and
$R$. It may be noted that two sets of independent special solutions $(S,R)$ are sufficient for
our purpose. Once these forms are determined the integral of motion
$I(t,x,\dot{x})$ can be deduced from the expression
\begin{eqnarray}
 I= r_1
  -r_2 -\int \left[R+\frac{d}{d\dot{x}} \left(r_1-r_2\right)\right]d\dot{x},
  \label{met13}
\end{eqnarray}
where
\begin{eqnarray}
r_1 = \int R(\phi+\dot{x}S)dt,\quad
r_2 =\int (RS+\frac{d}{dx}r_1) dx. \nonumber
\end{eqnarray}

Equation~(\ref{met13}) can be derived straightforwardly by
integrating Eq.~(\ref{met8}).
We solve Eq. (\ref{eq6}) through the extended PS procedure in the
following way.  For the given second-order ODE (\ref{eq6}), the first
integral $I$ should be either a time-independent or time dependent one.  In the
former case, it is a conservative system and we have $I_t=0$ and in the latter
case we have $I_t\ne 0$.  So, let us first consider the case $I_t=0$ and
determine the null forms and the corresponding integrating factors, and from
these we construct the integrals of motion and then we extend the analysis
to the case $I_t\ne 0$.

\section{Time independent integrals : Integrable parametric choices}
\label{sectid}
In this section, we identify a set of parametric choices of (\ref{eq6}) for which time
independent integrals exist.  For this purpose we first find the null forms and
integrating factors corresponding to equation (\ref{eq6}) and using these functions we
construct time independent integrals of (\ref{eq6}) through the relation (\ref{met13}).

\subsection{Null forms and integrating factors}
Since $I_t=0$, one can easily fix the null form $S$ from the first equation in (\ref{met8}) as
\begin{eqnarray}
S=\frac{-\phi}{\dot{x}}=\frac{(k_1\frac{\dot{x}^2}{x}+(k_2+k_3x)\dot{x}+k_4x^{3}+k_5x^{2}+k_6x)}{\dot{x}}.
\label{timeinds}
\end{eqnarray}
Of course one can easily check that the $S$ form given above satisfies Eq. (\ref{lin02}).

Substituting this form of $S$, given in (\ref{timeinds}), into (\ref{lin03}) we get
\begin{eqnarray}
\dot{x}R_x-(k_1\frac{\dot{x}^2}{x}+(k_2+k_3x)\dot{x}+k_4x^{3}+k_5x^{2}+k_6x) R_{\dot{x}}
=\bigg(\frac{k_1\dot{x}}{x}-\frac{k_4x^{3}+k_5x^{2}+k_6x}{\dot{x}}\bigg)R.
\label{timeindreq}
\end{eqnarray}
Since we are interested in time independent integrals we take $R_t=0$.
As we noted earlier any particular solution of the above equation along with the
null form $S$ is sufficient to construct an integral of motion.  To derive a particular
solution of (\ref{timeindreq}) we make an ansatz for $R$  of the form
\begin{eqnarray}
R=\frac{\dot{x}}{(A(x)+B(x)\dot{x}+C(x)\dot{x}^2)^r},\label{timeindr}
\end{eqnarray}
where $r$ is a constant and $A(x)$, $B(x)$ and $C(x)$ are arbitrary functions
of their argument.
The reason for choosing the above form of ansatz is as follows.  To deduce
the time independent first integral $I$ we assume a rational form for $I$, that is,
$I=\frac{f(x,\dot{x})}{g(x,\dot{x})}$, where $f$ and $g$ are arbitrary functions
of  $x$ and $\dot{x}$.  Using (\ref{met3}) we obtain
$S=I_x/I_{\dot{x}}=(f_xg-fg_x)/(f_{\dot{x}}g-fg_{\dot{x}})$.
and $R=I_{\dot{x}}=(f_{\dot{x}}g-fg_{\dot{x}})/g^2$ and from these two expressions
we find that the numerator of $R$ should be the
denominator of $S$ and so we fixed the numerator of the $R$ in Eq. (\ref{timeindr}) as $\dot{x}$ (see expression (\ref{timeinds}) for $S$), that is $R=\frac{\dot{x}}{h(x,\dot{x})}$, where $h$ is an arbitrary function.  However, it is difficult to proceed with this choice of $h$. So, we further assume that $h(x ,\dot{x})$ is a polynomial in $\dot{x}$.
To begin with we consider the case in which $h$ is a quadratic function in $\dot{x}$,
that is $h = A(x) + B(x) \dot{x}+C(x)\dot{x}^2$. Since $R$ is in rational form while taking differentiation or integration the form of the denominator remains the same but the power of the denominator increases or decreases by a unit order from that of the initial one. So, instead of considering
$h$ to be of the form $h = A(x) + B(x) \dot{x}+C(x)\dot{x}^2$, we consider a more general form
$h =( A(x) + B(x) \dot{x}+C(x)\dot{x}^2) ^r$, where $r$ is a constant to be determined.  We
note here that this form of ansatz played a crucial role in deducing the time independent
integrals of certain dissipative systems, see for example Ref. 20.

Substituting (\ref{timeindr}) into (\ref{timeindreq}) and
solving the resultant equation, we find that the solution exists only for certain
specific choices of $k_i$'s.  In the following we provide these parametric restrictions
with the resultant forms of R:
\begin{eqnarray}
&&\mbox{\emph{Case \bf (i) :}}\nonumber\\
&& k_5=k_4(3+2k_1)\rho_1,\,
k_6=k_4(1+k_1)(2+k_1)\rho_1^2,\,\,\rho_1=\frac{k_2}{k_3(1+k_1)},\,k_1\ne-1\nonumber\\
&&R=\left\{
\begin{array}{ll}
\displaystyle\frac{\dot{x}}
 {((2+k_1)\dot{x}+k_3x(\rho_1(2+k_1)+x))\dot{x}
+k_4x^2(\rho_1(2+k_1)+x)^2},&k_3^2<4k_4(2+k_1)\\
\vspace{-0.5cm}\\
\displaystyle \frac{\dot{x}x^{(2-r)k_1}}
{\left[k_3(r-1)(\rho_1(2+k_1)+x)x+(2+k_1)r\dot{x}\right]^{r}},&k_3^2\ge4k_4(2+k_1)\\
\end{array}
\right.
\end{eqnarray}
\begin{eqnarray}
&&\hspace{-0.5cm}\mbox{where }r=\frac{k_3^2\pm k_3\sqrt{k_3^2-4k_4(2+k_1)}}{2k_4(2+k_1)}.\nonumber\\
&&\hspace{-0.5cm}\mbox{For the choice }k_1=-1,\mbox{ we find }k_4=0,k_6=\frac{k_2k_5}{k_3},\,\,R=\frac{\dot{x}}{k_3x\dot{x}+k_5x^2}\\
&&\mbox{\emph{Case \bf (ii) :}}\nonumber\\
&&k_2=0,\,k_3=0, \,\,R=\dot{x}x^{2k_1}\\
&&\mbox{\emph{Case \bf (iii) :}}\nonumber\\
&&k_1=0,\,k_2=0,\,k_5=0,\nonumber\\
&&\displaystyle R=\left\{
\begin{array}{ll}
\displaystyle\frac{\dot{x}}{2k_4\dot{x}^2+k_3\dot{x}(k_4x^2+k_6)+(k_4x^2+k_6)^2},&k_3^2<8k_4\\
\displaystyle\frac{\dot{x}}{(\dot{x}+\frac{(r-1)}{2r}k_3x^2+\frac{rk_6}{k_3})^{r}},&k_3^2\ge8k_4,
\,\,r=\frac{k_3^2\pm k_3\sqrt{k_3^2-8k_4}}{4k_4}.
\end{array}
\right.
\end{eqnarray}

The question now remains as to whether the functions $S$ and $R$ given
above satisfy the third of the determining equations (vide Eq. (\ref{lin04})) or not.  One can
easily check that all the above three sets of functions $S$ and $R$ do indeed satisfy the equation
(\ref{lin04}). The parametric restrictions given above fix the equation of
motion (\ref{eq6}) to the following specific forms:
\begin{eqnarray}
&&\mbox{\underline{\emph{Case {\bf (i)} Eq. (\ref{tid5}):}} }\nonumber\\
&&\ddot{x}+k_1
\frac{\dot{x}^2}{x}+k_3(\rho_1(1+k_1)+ x)\dot{x} +
k_4x^3+k_4\rho_1(3+2k_1)x^2
+k_4(1+k_1)(2+k_1)\rho_1^2x=0,\nonumber\\
&&\mbox{where }\rho_1=\frac{k_2}{k_3(1+k_1)}\nonumber\\
&&\mbox{\underline{\emph {Case {\bf (ii)} Eq. (\ref{bern1}):}}}\nonumber\\
&&\ddot{x}+k_1\frac{\dot{x}^2}{x}+k_4 x^3+k_5 x^2+k_6 x=0.\,\nonumber%
\end{eqnarray}
\begin{eqnarray}
&&\hspace{-5cm}\mbox{\underline{\emph{Case {\bf (iii)} Eq. (\ref{tid4}):}}}\nonumber\\
&&\hspace{-2cm}\ddot{x}+k_3 x\dot{x}+k_4x^3+k_6x=0.\nonumber
\end{eqnarray}

In the next section, we construct time independent integrals
for the above equations.  We mention here that the last equation (\ref{tid4}) is nothing
but the modified Emden equation with additional linear force term which admits a conservative Hamiltonian
 description for all values of $k_3$, $k_4$ and $k_6$.  For the special choice,
 $k_4=\frac{k_3^2}{9}$, this equation has been shown to exhibit unusual phenomena like
amplitude independent frequency of oscillations and conservative Hamiltonian structure (for more details one may refer 21).

\subsection{Integrals of motion}
Having determined the explicit forms of $S$ and $R$, one can
proceed to construct the integrals of motion using the expression
(\ref{met13}) for the above cases. Substituting the corresponding forms of $S$ and $R$ into the general
form of the integral of motion (\ref {met13}) and evaluating the
resultant integrals, we obtain the following time independent
integrals for the above three cases:
\begin{eqnarray}
&&\hspace{-1cm}\mbox{\emph{Case \bf (i)}:  }\nonumber\\
&& I_1=\left\{
\begin{array}{ll}
\displaystyle\log\left[x^{2k_1}\{(2+k_1)\dot{x}^2+[(2+k_1)\rho_1+x]
[k_3x\dot{x}\right.\\
\left.\displaystyle\qquad+ k_4x^2((2+k_1)\rho_1+x)]\}\right]
-\frac{2k_3\tan^{-1}\left[\frac{\Phi(x,\dot{x})}{\Omega(x)}\right]}
{\sqrt{4k_4(2+k_1)-k_3^2}}
\,,&k_3^2<4k_4(2+k_1)\label{integral6}\\
\vspace{-0.3cm}\\
\displaystyle\frac{x^{k_1(2-r)}[k_3x((2+k_1)\rho_1+x)+(2+k_1)r\dot{x}]}
{\left\{k_3(r-1)x((2+k_1)\rho_1+x)+(2+k_1)r\dot{x}\right\}^{r-1}}
\,,\,&k_3^2>4k_4(2+k_1)\\\vspace{-0.3cm}\\
\displaystyle\log\bigg[\bigg(k_3x[(2+k_1)\rho_1+x]+2(2+k_1)\dot{x}\bigg)x^{k_1}\bigg]\\
\displaystyle\qquad-\frac{2(2+k_1)\dot{x}}
{k_3x((2+k_1)\rho_1+x)+2(2+k_1)\dot{x}},&k_3^2=4k_4(2+k_1)\\\vspace{-0.3cm}\\
k_2\frac{\dot{x}}{x}+k_2k_3x+\left(k_2^2+\frac{k_2k_5}{k_3}\right)
\log[x]-\frac{k_2k_5}{k_3}\log\left[k_2\dot{x}+\frac{k_2k_5}{k_3}\right],&k_1=-1
\end{array}
\right.\\\nonumber
&&\hspace{-1cm}\mbox{where }\Phi(x,\dot{x})=  k_3x^2+(2+k_1)
(k_3x\rho_1+2\dot{x}),\nonumber\\
&&\hspace{-1cm}\mbox{and }\Omega(x)=  \sqrt{4(2+k_1)k_4-k_3^2}
\bigg[(2+k_1)\rho_1+x\bigg]x.\nonumber\\
&&\hspace{-1cm}\mbox{\emph{Case \bf (ii)}:  }\nonumber\\
&&I_1=
\begin{array}{ll}
\displaystyle\bigg (\frac {\dot{x}^2} {x^2} +\frac {k_6}{1+k_1}
+\frac
{2(2+k_1)k_5x+(3+2k_1)k_4x^2}{(2+k_1)(2k_1+3)}\bigg)x^{2(1+k_1)},
 \label{integral8}
\end{array}
\end{eqnarray}
\begin{eqnarray}
&&\hspace{-1cm}\mbox{\emph{Case \bf (iii)}:  }\nonumber\\
&&I_1=\left \{
\begin{array}{ll}
\log[2k_4\dot{x}^2+k_3(k_4x^2+k_6)\dot{x}+(k_4x^2+k_6)^2]
\\
\qquad+\frac{2k_3\tan^{-1}\left[\frac{k_3\dot{x}+2(k_4x^2+k_6)}{\dot{x}
\sqrt{8k_4-k_3^2}}\right]}{\sqrt{8k_4-k_3^2}},&k_3^2<8k_4\\\vspace{-0.3cm}\\
\bigg(\dot{x}+\frac{(r-1)}{2r}k_3x^2+\frac{rk_6}{k_3}\bigg)^{-r}
\bigg\{\dot{x}\bigg(\dot{x}+\frac{k_3}{2}x^2+\frac{r^2k_6}{(r-1)k_3}\bigg)\\
\qquad+\frac{(r-1)}{r^2}\bigg(\frac{k_3}{2}x^2\label{tid3integral}
+\frac{r^2k_6}{(r-1)k_3}\bigg)^2\bigg\},\,&k_3^2>8k_4\\\vspace{-0.3cm}\\
\displaystyle\frac{4k_3\dot{x}}{k_3^2x^2+4k_3\dot{x}+8k_6}
-\log[k_3^2x^2+4k_3\dot{x}+8k_6],\,&k_3^2=8k_4,\\
\end{array}
\right.
\end{eqnarray}
Note that in the above the ranges of $x$ and $\dot{x}$ should be so restricted that no
multivaluedness occurs.
In a recent paper{\footnotesize$^{22}$} we have shown that the general equation of the form
\begin{eqnarray}
\ddot{x}+\frac{g'(x)}{g(x)}\dot{x}^2+\alpha \frac{f(x)}{g(x)}\dot{x} +\lambda \frac{f(x)}{g(x)^2}\int f(x)dx=0,\label{lienard}
\end{eqnarray}
 is related
to the damped harmonic oscillator equation
\begin{eqnarray}
y''+\alpha y'+\lambda y=0,\qquad \left('=\frac{d}{d\tau}\right)\label{damp}
\end{eqnarray}
through the nonlocal connection
\begin{eqnarray}
y=\int f(x)dx,\,\,\, d\tau=\frac{f(x)}{g(x)}dt.
\end{eqnarray}
By suitably choosing $f(x)$ and $g(x)$ one can show that the damped harmonic oscillator gets transformed to Eqs. (\ref{tid5}), (\ref{bern1}) and (\ref{tid4}).  By applying this transformation to the time independent integrals of motion of the damped harmonic oscillator one can also obtain the
above results.  However, it is not possible to obtain the time dependent integrals of motion of Eq. (\ref{eq6}) using the above form of nonlocal transformation.  We mention here that the time independent
integrals for the specific equation (\ref{tid4}) (Li\'enard type equation) can also be deduced by the procedure described in
Ref. 23.

\section{Time independent integrals of motion: integrability \& general solution}
\subsection{Hamiltonian description and integrability}
\label{hamdes}
In the previous subsection we showed that the equations (\ref{tid5}), (\ref{bern1}) and (\ref{tid4})
admit time independent integrals of motion.  Interestingly one can
interpret these integrals as time independent (but nonstandard) Hamiltonians for the respective systems and
they can be treated as conservative systems.  In the following we deduce the underlying Hamiltonian
structure for the Eqs. (\ref{tid5})-(\ref{tid4}) from the first integrals (\ref{integral6})-(\ref{tid3integral}). To do
so we assume a Hamiltonian of the form
\begin{eqnarray}
H(x,p)=I(x,\dot{x})=p\dot{x}-L(x,\dot{x}),\label{ham1}
\end{eqnarray}
where $L(x,\dot{x})$ is the Lagrangian and $p$ is the canonically conjugate
momentum. From (\ref{ham1}) we get
\begin{eqnarray}
\frac{\partial I}{\partial \dot{x}}=\frac{\partial p}{\partial
\dot{x}}\dot{x},\label{ham}
\end{eqnarray}
from which we identify
\begin{eqnarray}
p=\int\frac{I_{\dot{x}}}{\dot{x}}d\dot{x}.\label{momentum}
\end{eqnarray}

It is clear from (\ref{momentum}) that once $I$ is known $p$ can be
determined in terms of $\dot{x}$ and $x$ and inverting one can express $\dot{x}$ in terms of $x$ and $p$.  Substituting the expression for $\dot{x}$ in terms of $p$ into the expression for $I$ and from (\ref{ham1}) one can
deduce $L$.  Once $p$ and $L$ are known the same expression (\ref{ham1}) can
be utilized to derive $H$.  Using this procedure
one can deduce the Hamiltonian from the first integrals for all the three Cases (i)-(iii).
The Hamiltonians and the corresponding canonical
conjugate momenta read as follows:
\begin{eqnarray}
&&\hspace{-1cm}\mbox{{Case {\bf (i)}:} }\nonumber\\
&&H=\left\{
\begin{array}{ll}
\displaystyle\log\left[x^{k_1}(\rho_1(2+k_1)x+x^2)\sec\left[\frac{\Omega p}{2(2+k_1)}\right]\right]\\
\qquad-\frac{k_3p}{2(2+k_1)}(\rho_1(2+k_1)x+x^2),
&k_3^2<4k_4(2+k_1)\\\vspace{-0.6cm}\\
\mbox{where }\Omega(x)=  \sqrt{4(2+k_1)k_4-k_3^2}
\bigg[(2+k_1)\rho_1+x\bigg]x.\nonumber\\
\displaystyle\frac{(r-1)}{(r-2)}\left(px^{-k_1}\right)^{\frac{(r-2)}{(r-1)}}-\frac{k_3p(r-1)}{r(2+k_1)}\label{tidham1}
(\rho_1(2+k_1)x+x^2),
&k_3^2>4k_4(2+k_1)\\\vspace{-0.4cm}\\
\displaystyle\log\left[\frac{x^{k_1}}{p}\right]+\frac{k_3px}{2(2+k_1)}((2+k_1)\rho_1+x),&k_3^2=4k_4(2+k_1)\\\vspace{-0.4cm}\\
\end{array}
\right .
\end{eqnarray}
where
\begin{eqnarray}
&&\hspace{-1.5cm}\quad p=\left\{
\begin{array}{ll}
\frac{2(2+k_1)}{\Omega}\tan^{-1}\bigg[\frac{k_3x((2+k_1)\rho_1+x)+2(2+k_1)\dot{x}}
{\Omega}\bigg],
&\quad k_3^2<4k_4(2+k_1)
\\\vspace{-0.4cm}\\
\frac{x^{(2-r)k_1}}
{\left(\dot{x}+\frac{k_3(r-1)}{r(2+k_1)}((2+k_1)\rho_1x+x^2)\right)^{r-1}},
&\quad k_2^2\ge4k_4(2+k_1).\\\vspace{-0.4cm}\\
\vspace{-0.4cm}\\
\end{array}
\right.\\\nonumber\\
&&\hspace{-1cm}\mbox{{Case {\bf (ii)}:} }\nonumber\\
&& H=
x^{2(k_1+1)}\bigg (\frac{p^2}{4x^{2(2k_1+1)}}+\frac
{k_6}{k_1+1}+\frac{x(2(k_1+2)k_5+(2k_1+3)k_4x)}{2k_1^2+7k_1+6}\bigg ),\,p=2x^{2k_1} \dot{x}.\label{tid2ham}
\end{eqnarray}
\begin{eqnarray}
&&\hspace{-1cm}\mbox{{Case {\bf (iii)}:} }\nonumber\\
&& H=\left\{
\begin{array}{ll}
\displaystyle\frac{k_3}{4}p\left(x^2+\frac{k_6}{k_4}\right)
-\log\left[\left(x^2+\frac{k_6}{k_4}\right)\sec[\frac{\gamma p}{2}
(x^2+\frac{k_6}{k_4})]\right],\label{tid3ham}
&k_3^2<8k_4 \\\vspace{-0.4cm}\\
\displaystyle\frac{(r-1)}{(r-2)}
p^{\frac{(r-2)}{(r-1)}}-\frac{(r-1)}{2r}k_3px^2-\frac{rpk_6}{k_3},& k_3^2>8k_4\\\vspace{-0.4cm}\\
\displaystyle\log[p]-\frac{k_3}{2}p\left(\frac{x^2}{2}+\frac{4k_6}{k_3^2}\right),&k_3^2=8k_4\\
\end{array}
\right .\\\nonumber
&&\hspace{-1.5cm}\mbox{where}\\
&&\hspace{-1.5cm}\qquad\quad p=\left\{
\begin{array}{ll}
\frac{2}{\gamma( x^2+\frac{k_6}{k_4})}\tan^{-1}\left[\frac{4\dot{x}+k_3(x^2
+\frac{k_6}{k_4})}{2\gamma (x^2+\frac{k_6}{k_4})}\right],&k_3^2<8k_4,\,\,\gamma=\frac{1}{2}\sqrt{8k_4-k_3^2}
\\\vspace{-0.4cm}\\
\dot{x}+\frac{(r-1)}{2r}k_3x^2+\frac{rk_6}{k_3},\qquad\quad& k_3^2\ge8k_4.
\end{array}
\right .
\end{eqnarray}

One can check that in all the above cases the second order
equivalence of the Hamilton's equation of motion coincides exactly
with the  associated equations (\ref{tid5}), (\ref{bern1}) and
(\ref{tid4}) for the appropriate parametric choices.  The
existence of the time independent Hamiltonian for the Eqs.
 (\ref{tid5}), (\ref{bern1}) and (\ref{tid4}) assures us that they
are Liouville integrable{\footnotesize$^{4}$}.  However, our target is to
go beyond this statement and construct the solutions for these
equations.  On the other hand, we find that it is difficult to integrate the
Hamilton's equations of motion corresponding to the above Hamiltonians except for the second case (vide
Eq. (\ref{tid2ham})) whose solution can be deduced using the
standard methods.  In order to deduce the solutions we introduce
suitable canonical transformations so that the new Hamilton's
equations of motion can be integrated.  However, we are able to
construct the explicit solution of Eq. (\ref{tid5}) only for the
parametric choice $\rho_1=0$, making the resultant equation
equivalent to Eq. (\ref{eqiv}) whose solution is constructed in
the subsection \ref{caseivgensol}. Hence we do not discuss the
solution of case (i) separately here.
\subsection{General solution}
In this sub-section we discuss the method
of finding the general solution for the case (ii) and case (iii).

\n
{\bf{Case {\bf (ii)}:}}

We find that one can rewrite the first integral (\ref{integral8})
straightforwardly into the following quadrature
\begin{eqnarray}
t-t_0=\int\frac{dx}{\sqrt{I_1x^{-2k_1}
-\frac{k_4(3+2k_1)x^4+2k_5(2+k_1)x^3}{2k_1^2+7k_1+6}-\frac{k_6x^2}{1+k_1}}}.\label{solbern1}
\end{eqnarray}

On the other hand the
solution of (\ref{bern1}) can also be obtained by transforming it to the Bernoulli equation
through the transformation $\dot{x}=z(x)$, where $x$ is the new independent variable.  The solution of this reduced Bernoulli
type equation is again given in terms of quadratures{\footnotesize$^{2}$}.

\n
{\bf{Case {\bf (iii) a)}: Parametric choice $k_3^2<8k_4$}}

In order to obtain the solution for the under damped case ($k_3^2<8k_4$) we use the following
canonical transformation,
\begin{eqnarray}
x=-\sqrt{\frac{k_6}{k_4}}\tan\left[\sqrt{\frac{k_6}{k_4}}P\right],\quad
p=\frac{k_4U}{k_6}\cos^2\left[\sqrt{\frac{k_6}{k_4}}P\right],
\end{eqnarray}
and transform the Hamiltonian
$
H=\frac{k_3}{4}p\left(x^2+\frac{k_6}{k_4}\right)
-\log\left[\left(x^2+\frac{k_6}{k_4}\right)\sec[\frac{\gamma p}{2}
(x^2+\frac{k_6}{k_4})]\right]\nonumber
$
into
\begin{eqnarray}
H=\log\left[\frac{k_6}{k_4}\sec^2\left[\sqrt{\frac{k_6}{k_4}}P\right]
\sec\left[\frac{\gamma}{2}U\right]\right]-\frac{k_3}{4}U.
\label{tid3canham1}
\end{eqnarray}
The corresponding Hamilton equations of motion are
\begin{eqnarray}
\dot{U}=2\sqrt{\frac{k_6}{k_4}}\tan\left[\sqrt{\frac{k_6}{k_4}}P\right],\quad
\dot{P}=\frac{k_3}{4}-\frac{\gamma}{2}\tan\left[\frac{\gamma}{2}U\right].
\end{eqnarray}

From the first equation we can express $P$ in terms of $U$, that is,
$P=\sqrt{\frac{k_4}{k_6}}\tan^{-1}\left[\frac{\sqrt{k_4}\dot{U}}{2\sqrt{k_6}}\right]$ and
substituting this form of $P$ into (\ref{tid3canham1}) we obtain
\begin{eqnarray}
H=\log\left[\frac{(4k_6+k_4\dot{U}^2)}{4k_4}\sec\left[\frac{\gamma}{2}U\right]\right]
-\frac{k_3U}{4}\equiv E.
\label{eq59a}
\end{eqnarray}
Rewriting the last equation (\ref{eq59a}) for $\dot{U}$ as
\begin{eqnarray}
\dot{U}=2\sqrt{\exp\left[E+\frac{k_3}{4}U\right]\cos\left[\frac{\gamma}{2}U\right]
-\frac{k_6}{k_4}}
\end{eqnarray}
and integrating we arrive at the following quadrature
\begin{eqnarray}
t-t_0=\int\frac{dU}
{2\sqrt{\exp\left[E+\frac{k_3}{4}U\right]\cos\left[\frac{\gamma}{2}U\right]
-\frac{k_6}{k_4}}}.
\end{eqnarray}
\n
{\bf{Case {\bf (iii) b)}: Parametric choice $k_3^2>8k_4$}}

In order to obtain the solution for the over damped case ($k_3^2>8k_4$), we use the following canonical transformation
$x=\frac{P}{U},\quad p=\frac{U^2}{2}$ and obtain a new Hamiltonian of the form
\begin{eqnarray}
H=\sigma_1U^{\frac{2}{r_{12}}}
-\left(\frac{k_3(r-1)P^2}{4r}+\frac{k_6rU^2}{2k_3}\right),
\end{eqnarray}
where $\sigma_1=\frac{r_{12}}{2^{\frac{1}{r_{12}}}}$.
The canonical equations in the new variable assume the form
\begin{eqnarray}
\dot{U}=\frac{k_3P(1-r)}{2r},\quad \dot{P}=\frac{k_6rU}{k_3}-\frac{2U^{\frac{2}{r_{12}}}}
{r_{12}U}.
\end{eqnarray}
From the first relation one can express $P$ in terms of $\dot{U}$, that is
$P=\frac{2r\dot{U}}{k_3(1-r)}$.  Substituting the latter in the Hamiltonian we get
\begin{eqnarray}
H\equiv E=\sigma_1U^{\frac{2}{r_{12}}}-\frac{k_6r}{2k_3}U^2
+\frac{r\dot{U}^2}{k_3(1-r)}
\end{eqnarray}
which in turn leads us to the quadrature of the form
\begin{eqnarray}
t-t_0=\left(\frac{k_3(1-r)}{r}\right)^{\frac{1}{2}}\int
\frac{dU}{\sqrt{E-\sigma_1U^{\frac{2}{r_{12}}}+\frac{k_6rU^2}{2k_3}}}
\end{eqnarray}
\n
{\bf{Case {\bf (iii) c)}: Parametric choice $k_3^2=8k_4$}}

Finally, now we focus our attention on the critically damped case ($k_3^2=8k_4$).
Using the same canonical transformation $x=\frac{P}{U}$, $p=\frac{U^2}{2}$ we rewrite the
underlying Hamiltonian (\ref{tid3ham}) as
\begin{eqnarray}
H=2\log[U]-\frac{k_3}{4}\left(\frac{P^2}{2}+k_6U^2\right).
\end{eqnarray}
The corresponding canonical equations turn out to be
\begin{eqnarray}
\dot{U}=\frac{-k_3P}{4},\quad \dot{P}=\frac{k_3k_6}{2}-\frac{2}{U}.
\end{eqnarray}
Substituting $P=-\frac{4\dot{U}}{k_3}$ in the Hamiltonian, we get
\begin{eqnarray}
H\equiv E=2\log[U]-\frac{1}{4}k_3k_4U^2-\frac{2}{k_3}\dot{U}^2
\end{eqnarray}
Rearranging we get,
\begin{eqnarray}
\dot{U}=
\left(\frac{8k_3\log[U]-k_3^2k_6U^2-4Ek_3}{2}\right)^{\frac{1}{2}},
\end{eqnarray}
which upon integrating reduces to the following form of quadrature
\begin{eqnarray}
t-t_0=\sqrt{2}\int\frac{dU}{\sqrt{8k_3\log[U]-k_3^2k_6U^2-4Ek_3}}.
\end{eqnarray}

Summarizing the results, we find that the nonlinear equations (\ref{tid5}), (\ref{bern1}) and (\ref{tid4}) admit time
independent Hamiltonians for all values of the system parameters and can be classified
as integrable ones in the Liouville sense.  While constructing the solution for these
three equations
we find that the nonlinear system (\ref{bern1}) can be transformed into Abel equation which
in turn can be integrated into a quadrature. Solution of Eq. (\ref{tid4}) is
given in terms of quadrature by applying suitable canonical transformation to its Hamiltonian.  By using again canonical transformations the Hamiltonian corresponding to Eq. (\ref{tid5}) can be transformed and the canonical equations corresponding to this transformed Hamiltonian can be integrated and the solution is given in terms
of quadratures for the parametric choice $\rho_1=0$ (see Sec. \ref{caseivgensol}).
\section{Time dependent integrals $(I_t\ne0)$}
\label{sectd}
In this section, we explore the parametric choices for which (\ref{eq6}) admits
time dependent integrals.  The underlying procedure is same as that of the time
independent integral case but involves somewhat lengthy calculations.  As a first step
in this process we derive the null forms and integrating factors corresponding to Eq. (\ref{eq6})
by solving the determining Eqs. (\ref{lin02})-(\ref{lin04}).

\subsection{Null forms and integrating factors}

\n In the previous section we considered the case $I_t=0$.  As a
consequence $S$ turns out to be $\frac{-\phi}{\dot{x}}$.  However, in
the case $I_t\ne0$, the function $S$ has to be determined from
Eq. (\ref{lin02}), that is
\begin{eqnarray}
&&S_t+\dot{x}S_x-\bigg(\frac{k_1
\dot{x}^2}{x}+(k_2+k_3x)\dot{x}+k_4x^3+k_5x^2+k_6x\bigg)S_{\dot{x}}=-\frac{k_1\dot{x}^2}{x^2}\nonumber
\\&&\qquad+k_3\dot{x}+3k_4x^2+2k_5x+k_6-S\left(\frac{2k_1\dot{x}}{x}+(k_2+k_3x)\right)+S^2.
\label{sequation}
\end{eqnarray}
Since it is too difficult to solve Eq. (\ref{sequation}) for its general
solution, we seek a particular solution for S, which is sufficient for our
purpose.  The time independent integral case clearly indicates that
$S$ should be in a rational form.  To begin with one may consider $S=\frac{f(t,x,\dot{x})}
{g(t,x,\dot{x})}$, where $f$ and $g$ are arbitrary functions.  However, it is difficult to solve Eq. (\ref{sequation}) with this
form of $S$.  So one may assume that $f$ and $g$ are simple polynomials in $\dot{x}$
with coefficients which are arbitrary functions of $t$ and $x$.   We seek a simple rational
expression for $S$ in the form
\begin{eqnarray}
S=\frac{a(x,t)+b(x,t)\dot{x}+c(x,t)\dot{x}^2}{d(x,t)+e(x,t)\dot{x}+f(x,t)\dot{x}^2},
\label{sform}
\end{eqnarray}
where $a$, $b$, $c$, $d$, $e$ and $f$ are arbitrary functions of
$x$ and $t$ which are to be determined.  One may also consider a
cubic polynomial in $\dot{x}$ both in the numerator and in the
denominator.  However, the resultant analysis did not yield any
new result. So we confine our presentation here to the form (\ref{sform}) only.  Substituting (\ref{sform}) into (\ref{sequation}) and
equating the coefficients of different powers of $\dot{x}$ to
zero, we get
\begin{eqnarray}
&&-a^2+a_td-ad_t+adk_2-d^2k_6+adk_3x-2d^2k_5x-bdk_6x+aek_6x-3d^2k_4x^2
\nonumber\\
&&\qquad\qquad\qquad\qquad\qquad\qquad\qquad-bdk_5x^2+aek_5x^2-bdk_4x^3+aek_4x^3=0,\nonumber\\
&&a_xd-2ab-ad_x+b_td-bd_t+a_te-ae_t+2aek_2-d^2k_3-2dek_6+\frac{2adk_1}{x}
+2aek_3x
\nonumber\\
&&\qquad\qquad\qquad\qquad\qquad\qquad-4dek_5x-2cdk_6x-6dek_4x^2-2cdk_5x^2-2cdk_4x^3=0,\nonumber
\end{eqnarray}
\begin{eqnarray}
&&b_xd-b^2-2ac-bd_x+c_td-cd_t+a_xe-ae_x+b_te-be_t-cdk_2
+bek_2-2dek_3-e^2k_6
\nonumber\\
&&\qquad\qquad\qquad\qquad+\frac{d^2k_1}{x^2}+\frac{bdk_1}{x}+\frac{3aek_1}{x}-cdk_3x
+bek_3x-2e^2k_5x-cek_6x\nonumber\\
&&\qquad\quad\qquad\qquad\qquad\qquad\qquad\qquad\qquad-3e^2k_4x^2-cek_5x^2-cek_4x^3=0,\nonumber\\
&&c_xd-cd_x-2bc+b_xe-be_x+c_te-ce_t-e^2k_3+\frac{2dek_1}{x^2}+\frac{2bek_1}{x}=0,\nonumber\\
&&c_xe-c^2-ce_x+\frac{e^2k_1}{x^2}+\frac{cek_1}{x}=0.
\label{detereq}
\end{eqnarray}

We obtain four parametric choices for which we are able to find
nontrivial particular solutions of the above set of coupled
partial differential equations. We present the parametric choices
and the null forms $S$ in Table-I. We find that the null forms $S$
obtained are incidently independent of time.

Next, the above identified forms of $S$ along with their corresponding parametric
restrictions are substituted in the determining equation (\ref{lin03}) for $R$. To solve
the resultant equation for $R$ we make use of the ansatz
\begin{eqnarray}
 R=\frac{F(t)S_d }{(A(x)+B(x)\dot{x}+C(x)\dot{x}^2)^r},\label{ransatz}
\end{eqnarray}
 where $S_d$ is the denominator of $S$. We demand
the above form due to the following reason.  To deduce
the first integral $I$ we assume a rational form for $I$, that is,
$I=\frac{f(t,x,\dot{x})}{g(t,x,\dot{x})}$, where $f$ and $g$ are
arbitrary functions of $t$, $x$ and $\dot{x}$.  From (\ref{met8}), we know that
$S=I_x/I_{\dot{x}}=(f_xg-fg_x)/(f_{\dot{x}}g-fg_{\dot{x}})$ and as
we find the deduced $S$ forms are independent of time, the
numerator and denominator of $S$ should share a common factor
which is a function of time alone, that is, the numerator and
denominator of $S$ should be of the form
$(f_xg-fg_x)=F(t)f_1(x,\dot{x})$ and
$(f_{\dot{x}}g-fg_{\dot{x}})=F(t)g_1(x,\dot{x})$, respectively.
Moreover, from the relation
$R=I_{\dot{x}}=(f_{\dot{x}}g-fg_{\dot{x}})/g^2$, we find that the
numerator of $R$ should be the denominator of $S$.

On solving the resultant equation,
we obtain the integrating factor $R$.  We present the forms of $R$
along with the $S$ forms in Table 1.  Once $S$ and $R$ are determined then one
has to verify the compatibility of this set $(S,R)$ with the extra
constraint Eq. (\ref{lin04}). Having verified the compatibility of
$(S,R)$, we substitute $S$ and $R$ into Eq. (\ref{met13}) and
construct the associated integral of motion.  In this way we identify
four sets of integrable parametric choices.  We present the explicit form
of these integrals of motion in Sec \ref{application}.
\begin{sidewaystable}

\caption{Parametric restrictions, null forms $S$ and integrating
factors $R$ of\\
 $\ddot{x}+k_1 \frac{\dot{x}^2}{x}+(k_2+k_3
x)\dot{x}+k_4 x^3+k_5 x^2+k_6 x=0$}

\noindent\begin{tabular}{|l|l|l|l|}
\hline Case & Parametric restriction &\qquad\qquad\qquad Form of $S$  &\qquad\qquad\qquad Form of $R$ \\
\hline & & & \\ (i) &$k_4=\frac{(1+k_1) k_3^2}{(3+2 k_1)^2},$
 & &   \\
 & $k_5=\frac{k_2 k_3}{3+2k_1}.$&$\qquad\qquad\quad\frac{k_3 x}{(3+2 k_1)}-\frac{\dot{x} }{x}$ &$\qquad\frac{\pm e^{\pm\omega t}x}
{\bigg(\dot{x}+\frac{1}{2}\frac{k_2\pm\omega}{1+k_1}x+\frac{k_3}{3+2k_1}x^2\bigg)^2}$\\
 &$k_1$, $k_2$, $k_3$ arbitrary & &$\qquad\omega=\sqrt{k_2^2-4(1+k_1)k_6}$ \\
\hline (ii) &$k_5=\frac{k_3(k_2\pm\omega)}{2 (2+k_1)},$
& & \\
&$k_4=0.$ &$\qquad\quad\frac{1}{2}(k_2\mp\omega+2 k_3 x)
+k_1\frac{\dot{x}}{x}$ &$\qquad\qquad e^{\frac{1}{2}(k_2\pm\omega)t}
x^{k_1}$\\
& $k_1$, $k_2$, $k_3$ arbitrary&  & \\
\hline (iii) &$k_6=\frac{2(3+2 k_1)k_2^2}{(5+4 k_1)^2},$
&&   \\
&$k_3=0,k_4=0.$ &$\frac{\frac{4(1+k_1)k_2^2}{5+4k_1} x^2+(5+4k_1)k_5 x^3
+2(1+2k_1)k_2 x\dot{x}+k_1(5+4 k_1)\dot{x}^2)}{2 k_2x^2+(5+4 k_1)x\dot{x}}$ &
$e^{\frac{2(3+2k_1)k_2 t}{5+4 k_1}}x^{2 k_1}\bigg(2 k_2 x+(5+4k_1)\dot{x}\bigg)$\\
& $k_1$, $k_2$ arbitrary& & \\
& & & \\
\hline
\end{tabular}
\end{sidewaystable}
\begin{sidewaystable}
\noindent\begin{tabular}{|l|l|l|l|}
\hline Case \,\, & Parametric restriction &\qquad\qquad\qquad Form of $S$  & \qquad\qquad\quad\qquad\qquad
Form of $R$  \\
\hline & & & \\
(iva) &$k_5=\frac{k_2 k_3}{3 + 2k_1},$
& &
$\bigg(k_2x+(3+2k_1)\dot{x}\bigg)
\bigg[(2+k_1)k_2^2x^2+(3+2k_1)k_2x(k_3x^2+2(2+k_1)\dot{x})$ \\
 &$k_6=\frac{(2+k_1)k_2^2}{(3+2k_1)^2}$. &$k_1\frac{\dot{x}}{x}+k_3x+\frac{(3+2k_1)k_4x^3}{(3+2k_1)\dot{x}+k_2x}+
\frac{(1+k_1)k_2}{3+2k_1}$  &$+(3+2k_1)^2
(k_4x^4+\dot{x}(k_3x^2+(2+k_1)\dot{x}))\bigg]^{-1}$\\
 &$k_1$, $k_2$, $k_3$, $k_4$ arbitrary &   &
 \\
&&&\\
\hline
(ivb) &$k_5=\frac{k_2k_3}{3+2k_1}$,&&\\
&$k_6=\frac{(2+k_1)k_2^2}{(3+2k_1)^2},$ & $k_1\frac{\dot{x}}{x}+k_3x+\frac{(3+2k_1)k_4x^3}{(3+2k_1)\dot{x}+k_2x}$  &
\\
&$k_4=\frac{(r-1)k_3^2}{(2+k_1)r^2}. $ &  &$\qquad
\frac{e^{\frac{(2-r)(2+k_1)k_2 t}{3+2k_1}}(k_2x+(3+2k_1)\dot{x})
x^{k_1(2-r)}}{\bigg((2+k_1)r\left(\frac{k_2x}{3+2k_1}+\dot{x}\right)+k_3(r-1)x^2\bigg)^r}$ \\
&$k_1$, $k_2$, $k_3$ arbitrary &  & \\
\hline
\end{tabular}
\end{sidewaystable}

\subsection{Integrals of motion and general solution}
\label{application}
\n {\bf Case (i)} $k_1,\,k_2,\,k_3,\,k_6$  : arbitrary, $\displaystyle k_4=\frac{(1+k_1) k_3^2}{(3+2
k_1)^2},\,k_5=\frac{k_2 k_3}{3+2k_1}$

The parametric restriction given above fixes the equation of
motion (\ref{eq6}) as
\begin{eqnarray}
\ddot{x}+k_1\frac{\dot{x}^2}{x}+(k_2+k_3x)\dot{x}+\frac{k_3^2(1+k_1)}{(3+2k_1)^2}x^3+\frac{k_2k_3}{(3+2k_1)}x^2+k_6x=0.\nonumber
\qquad\qquad\qquad(\ref{caseia,b})
\end{eqnarray}
Equation (\ref{caseia,b}) reduces to the generalized modified Emden
equation for the parametric choice $k_1=0$ whose integrability has been
 studied in detail in Ref. 5. The integral of motion
associated with Eq. (\ref{caseia,b}) for $k_1\ne0$ turns out to be
\begin{eqnarray}
&&\mbox{\bf(ia)}\,\,\,\,\,I_1=e^{\pm \omega t}
\bigg(\frac{\dot{x}+
\frac{(k_2\mp\omega)}{2(1+k_1)}x+\rho_2x^2}{\dot{x}
+\frac{(k_2\pm\omega)}{2(1+k_1)}x+\rho_2x^2}\bigg),\,\,\,
\,\qquad\qquad\qquad\qquad\,\,\,\, k_1\ne-1,\omega\ne0\label{integral1}\\\vspace {-0.5 cm}\nonumber\\
&&\mbox{\bf(ib)}\,\,\,\,\,I_1=e^{k_2 t}\bigg (
\frac{k_2(\dot{x}+k_3x^2)+k_6x}{x}\bigg ),\,\,\,\,\,\,\,\,\,\,\qquad\qquad\qquad\qquad
k_1=-1\label{integral2}\\\vspace {-0.5 cm}\nonumber\\
&&\mbox{\bf(ic)}\,\,\,\,\,I_1=t-\frac{2(3+2k_1)x}{(3+2k_1)k_2x+2(1+k_1)(k_3x^2+(3+2k_1)\dot{x})},\,\,\,\,\,\omega=0\label{int3}
\end{eqnarray}
where $\rho_2=\frac{k_3}{3+2k_1}$, $\omega=\sqrt{k_2^2-4(1+k_1)k_6}$.

 Rewriting (\ref{integral1}) - (\ref{int3}) as first order ODEs we find that
the resultant equations are of Bernoulli type
which can be solved using the standard method{\footnotesize$^{2}$}.  The general solution to (\ref{caseia,b}) in each of these cases
 turns out to be
\begin{eqnarray}
&&\hspace{-0.8cm}\mbox{\bf (ia) }x(t)=\left\{c_2e^{\frac{c_2t}{2}}
\left[\frac{I_1-e^{\pm\omega t}}{I_1}\right]^\frac{\pm c_2}
{2\omega}(e^{\pm\omega t}-I_1)^\frac{\pm c_1}{2\omega}I_2
-2\rho_2\left[\frac{I_1-e^{\pm\omega t}}{I_1}\right]^
\frac{\pm c_1}{2\omega}(e^{\pm\omega t}-I_1)^{
\frac{\pm c_2}{2\omega}}\right.\nonumber\\
&&\qquad\quad\quad\left.\times F\left[\frac{\pm(c_1-c_2)}{2\omega},
\frac{\mp c_2}{2\omega},1\mp\frac{c_2}{2\omega},\frac{e^{\pm\omega t}}
{I_1}\right]\right\}^{-1}c_2\left\{\frac{-(I_1-e^{\pm\omega t})^2}
{I_1}\right\}^{\frac{\pm c_2}{2\omega}}\hspace{-0.5cm},
k_1\ne-1\label{solia}\\\vspace {-0.5 cm}\nonumber\\
&&\hspace{-0.8cm}\mbox{\bf (ib)}\,x(t)=(I_2 e^{\Phi}+k_3e^\Phi\int e^{-\Phi}dt)^{-1},
\hspace{6.5cm}\,k_1=-1\\\vspace {-0.5 cm}\nonumber\\
&&\hspace{-0.8cm}\mbox{\bf (ic)}\,x(t)=\frac{k_2(3+2k_1)\left(-c_3\right)^{\frac{1}{1+k_1}}
\mbox{exp}\left[\frac{c_3(I_1-t)}{2}\right]}
{(I_1-t)^{\frac{1}{1+k_1}}}\left(2^{\frac{2+k_1}{1+k_1}}(1+k_1)k_3
\Gamma\left[\frac{2+k_1}{1+k_1},\frac{c_3(t-I_1)}{2}\right]\right.
\nonumber\\
&&\qquad\qquad\quad\left.-I_2(3+2k_1)k_2(
-c_3)^{\frac{1}{1+k_1}}\right)^{-1},\,\omega=0
\end{eqnarray}
where $F$ is the hypergeometric function, $\Gamma$ is the gamma function, $c_1=\frac{k_2\mp\omega}{1+k_1}$,
$c_2=\frac{k_2\pm\omega}{1+k_1}$, $c_3=\frac{k_2}{1+k_1}$, $\Phi=\frac{k_2k_6t+e^{-k_2t}I_1}{k_2^{2}}$, $I_1$ and
$I_2$ are the integration constants.\\
{\bf{\bf Case (ii)} $k_1$, $k_2$, $k_3$ }: arbitrary,
$k_5=\frac{k_3(k_2\pm\omega)}{2(2+k_1)},\,k_4=0,\,
\omega=\sqrt{k_2^2-4(1+k_1)k_6}
$

The parametric restriction given above fixes the equation of
motion (\ref{eq6}) as
\begin{eqnarray}
&&\ddot{x}+k_1 \frac{\dot{x}^2}{x}+(k_2+k_3x)
\dot{x}+\frac{k_3(k_2\pm\omega)}{2(2+k_1)} x^2+k_6
x=0.\hspace{3cm}(\ref{eq87})\nonumber
\end{eqnarray}
The system (\ref{eq87}) possesses a
first integral of the form
\begin{eqnarray}
&&I_1=e^{\frac{1}{2}(k_2\pm\omega)t}x^{k_1}\bigg(\dot{x}+\frac{k_2\mp\omega}{2(1+k_1)}x+\rho_3x^2\bigg),\,\,k_1\ne-1,
\label{riccati1}
\end{eqnarray}
where $\rho_3=\frac{k_3}{(2+k_1)}$.

For the parametric choice $k_1=-1$, we obtain the same first integral as we have obtained in Case (ib) above.
On the other hand for the choice $k_1=-2$, Eq. (\ref{riccati1}) gets reduced
 to the Riccati equation which in turn  can
be integrated by the standard methods and the solution can be obtained of the form{\footnotesize$^{2}$}
\begin{eqnarray}
x(t)=\frac{2c_1e^{c_1t}}{I_1+2I_2c_1e^{2c_1t}-2\rho_3e^{c_1t}},
\end{eqnarray}
where $c_1=\frac{1}{2}(k_2\pm\omega)$, $I_1$ and $I_2$ are the integration constants.

For other choices of $k_1$ one is able to integrate the first integral (\ref{riccati1}) only by
imposing additional parametric restriction.  For example, choosing
$\displaystyle k_6=\frac{(2+k_1)k_2^2}{(3+2k_1)^2}$, and $\displaystyle k_5=\frac{k_3k_2(1+k_1)}{(3+2k_1)}$, one
is able to transform the time dependent integral of motion (\ref{riccati1}) into the following
time independent integral of motion
\begin{eqnarray}
I_1=w'+\frac{k_3(1+k_1)}{2+k_1}w^{\frac{2+k_1}{1+k_1}},\quad '=\frac{d}{dz},
\label{transintcii}
\end{eqnarray}
where $w$ and $z$ are the new dependent and independent variables, respectively and are given by
\begin{eqnarray}
w=\displaystyle e^{\frac{(1+k_1)k_2t}{3+2k_1}}x^{1+k_1},\qquad\,z=-\frac{(3+2k_1)e^{-\frac{k_2t}{3+2k_1}}}{k_2}.
\label{transformii}
\end{eqnarray}
The additional parametric restrictions fix Eq. (\ref{eq87}) to the specific form
\begin{eqnarray}
&&\ddot{x}+k_1 \frac{\dot{x}^2}{x}+\left(k_2+k_3x\right)\dot{x}+\frac{k_3k_2(1+k_1)}{(3+2k_1)} x^2
+\frac{(2+k_1)k_2^2}{(3+2k_1)^2} x=0,\hspace{1cm}(\ref{td2-1})\nonumber
\end{eqnarray}
Integrating (\ref{transintcii}) we get,
\begin{eqnarray}
z-z_0&=&\int\frac{dw}{I_1-\frac{k_3(1+k_1)}{2+k_1}w^{\frac{2+k_1}{1+k_1}}},\nonumber\\
&=&
wF\bigg[\frac{1+k_1}{2+k_1},1,\frac{3+2k_1}{2+k_1},
\frac{\hat{k}w^{\frac{2+k_1}{1+k_1}}}{I_1}\bigg]I_1^{-1},\label{solii}
\end{eqnarray}
where $F$ is the hypergeometric function{\footnotesize$^{24}$} and
$\displaystyle\hat{k}=\frac{k_3(1+k_1)}{2+k_1}$.
We also mention here that Eq. (\ref{td2-1})
contains
several sub-cases which are already known to be integrable, see for example Ref. 1.
\\
\noindent{\bf Case (iii)} $k_1$, $k_2$, $k_5$ : arbitrary,
$ k_3=0,\,k_4=0,\,k_6=\frac{2(3+2 k_1)k_2^2}{(5+4 k_1)^2}$

The parametric restriction given above fix the equation of motion (\ref{eq6}) as
\begin{eqnarray}
\ddot{x}+k_1 \frac{\dot{x}^2}{x}+\rho_5(5+4k_1)\dot{x}+k_5
x^2+2(3+2 k_1)\rho_5^2
x=0,\nonumber\qquad\qquad\qquad\qquad\qquad\quad(\ref{td3eq})
\end{eqnarray}
where $\displaystyle \rho_5=\frac{k_2}{(5+4k_1)}$.
The first integral for this equation is
\begin{eqnarray}
&\,I_1=x^{2k_1}\bigg(\dot{x}^2+4 \rho_5^2 x^2
+\frac{2 k_5 x^3}{3+2k_1}+4 \rho_5x\dot{x} \bigg)e^{2(3+2k_1) \rho_5t}.\label{integral3}
\end{eqnarray}
This time independent integral can be transformed into a time independent integral
by introducing a transformation of the form
\begin{eqnarray}
w=\frac{1}{\sqrt{2}}x^{1+k_1}
e^{2(1+k_1)\rho_5t},\,
z=-\frac{e^{-\rho_5t}}{\rho_5}.
\end{eqnarray}
where $w$ and $z$ are new dependent and independent variables, respectively.
Rewriting the integral of motion in terms of the new variables, we obtain
\begin{eqnarray}
I_1=w'^{2}+2^{\left(\frac{2k_1+3}{2(1+k_1)}\right)}\frac{k_5(1+k_1)^2}
{(3+2k_1)}w^{\frac{3+2k_1}{(1+k_1)}}.
\label{eq90a}
\end{eqnarray}
Eq. (\ref{eq90a})
can be integrated further and one can obtain the general solution as
\begin{eqnarray}
z-z_0=w\sqrt{\frac{I_1}{I_1-\hat{k} w^{\frac{3+2 k_1}{1+k_1}}}}
F\bigg[\frac{1+k_1}{3+2
k_1},\frac{1}{2},\frac{4+3k_1}{3+2k_1},\frac{\hat{k}w^{\frac{3+2k_1}{1+k_1}}}{I_1}
\bigg]
\end{eqnarray}
where $F$ is the hypergeometric function{\footnotesize$^{24}$} and
$\hat{k} = 2^{\frac{(3+2k_1)}{2(1+k_1)}}\frac{k_5(1+k_1)^2} {(3+2k_1)}$.

\vskip 6pt
\n {\bf Case (iv) } $k_1$, $k_2$, $k_3$ : arbitrary,
$ k_5=\frac{k_2 k_3}{3 + 2k_1},\,k_6=\frac{(2+k_1)k_2^2}{(3+2k_1)^2}.
$

The equation of motion in this case turns out to be
\begin{eqnarray}
&&\ddot{x}+k_1\frac{\dot{x}^2}{x}+(k_2+k_3x)\dot{x}+k_4x^3+\frac{
k_2k_3}{(3+2k_1)} x^2+\frac{k_2^2(2+k_1)}{(3+2k_1)^2}x=0.\nonumber\qquad\qquad\qquad
(\ref{eqiv})
\end{eqnarray}
The first integral reads
\begin{eqnarray}
&&\hspace{-1cm}\mbox{\bf(iva)}\,\,I_1=
\log\bigg[x^{2k_1}\bigg((2+k_1)\left(\dot{x}+\rho_6x\right)^2+k_3x^2(\dot{x}+\rho_6x)+k_4x^4
 \bigg)\bigg]\nonumber\\
&&\qquad\qquad-\frac{2k_3
\tan^{-1}\left(\frac{2(2+k_1)(\dot{x}+\rho_6x)+k_3x^2}
{x^2\sqrt{4(2+k_1)k_4-k_3^2}}\right)}
{\sqrt{4(2+k_1)k_4-k_3^2}}+2(2+k_1)\rho_6t,\quad4k_4(2+k_1)>k_3^2\nonumber\\\vspace {-0.5 cm}\nonumber\\
&&\hspace{-1cm}\mbox{\bf(ivb)}\,\,I_1=\frac{x^{k_1(2-r)}\left(k_3x^2+(2+k_1)r(\dot{x}+x\rho_6)\right)e^{(2+k_1)(2-r)\rho_6 t}}
{\left(k_3(r-1)x^2+(2+k_1)r(\dot{x}+x\rho_6)\right)^{r-1}},\qquad
4k_4(2+k_1)<k_3^2\nonumber\\
&&\hspace{-1cm}\mbox{\bf(ivc)}\,\, I_1=(2+k_1)\rho_6t+\log\left[k_3x^{2+k_1}+
2(2+k_1)x^{k_1}(\dot{x}+\rho_6x)\right]
\nonumber\\
&&\hspace{4cm}-
\frac{2(2+k_1)(\dot{x}+\rho_6x)}{k_3x^2+2(2+k_1)(\dot{x}+x\rho_6)},\qquad\quad\,
\hspace{0.3cm}4k_4(2+k_1)=k_3^2,\nonumber
\end{eqnarray}
where $\rho_6=\frac{k_2}{(3+2k_1)}$, $r=\frac{k_3^2\pm k_3\sqrt{k_3^2-4k_4(2+k_1)}}{2k_4(2+k_1)}$.

In the above forms of $I_1$, we now
introduce the following transformation,
\begin{eqnarray}
w=xe^{\rho_6t},\quad\,z=-\frac{e^{-\rho_6t}}{\rho_6},
\end{eqnarray}
so that in the new variables the integrals of motion read,
\begin{eqnarray}
&&\hspace{-2cm}\mbox{\bf(iva)}\,\,
 I_1=
\displaystyle\log\left[w^{2k_1}\{(2+k_1)w'^2+
[k_3w^2w'
+ k_4w^4]\}\right]-\frac{2k_3\tan^{-1}\left[\frac{ k_3w^2+2(2+k_1)
w'}{w^2\sqrt{4(2+k_1)k_4-k_3^2}}\right]}
{\sqrt{4k_4(2+k_1)-k_3^2}},
\\
&&\hspace{-2cm}\mbox{\bf(ivb)}\,\,I_1=
\displaystyle\frac{w^{k_1(2-r)}[k_3w^2+(2+k_1)rw']}
{\left\{k_3(r-1)w^2+(2+k_1)rw'\right\}^{r-1}},
\\
&&\hspace{-2cm}\mbox{\bf(ivc)}\,\,I_1=
\displaystyle\log\bigg[\bigg(k_3w^2+2(2+k_1)w'\bigg)w^{k_1}\bigg]-\frac{2(2+k_1)w'}
{k_3w^2+2(2+k_1)w'}.
\end{eqnarray}
In terms of the new variables $w$ and $z$ Eq. (\ref{eqiv}) reduces to the form
\begin{eqnarray}
w''+k_1\frac{w'^2}{w}+k_3ww'+k_4w^3=0\label{rho0}
\end{eqnarray}
which is equivalent to Eq. (\ref{tid5}) with $k_2=0$.
We find that it is very difficult to integrate the resultant integrals even after removing the time dependent factors.  To
establish the integrability of Eq. (\ref{eqiv}) we transform these time independent integrals
into time independent Hamiltonian and thereby establish the Liouville integrability of Eq. (\ref{eqiv}).

By following the procedure given in Sec. \ref{hamdes} we obtain the following Hamiltonian for the above cases, namely
\begin{eqnarray}
&&\hspace{-1.1cm}\mbox{\bf(iva)}\,\,H
=\displaystyle\log\left[w^{2+k_1}\sec\left[\frac{w^2\sqrt{4(2+k_1)k_4-k_3^2} p}{2(2+k_1)}\right]\right]
-\frac{k_3pw^2}{2(2+k_1)},\,\,4k_4(2+k_1)>k_3^2\label{ivham1}\\
&&\hspace{-1.1cm}\mbox{\bf(ivb)}\,\,H
=\displaystyle\frac{(r-1)}{(r-2)}\left(pw^{-k_1}\right)^{\frac{(r-2)}{(r-1)}}-\frac{k_3p(r-1)}{r(2+k_1)}
w^2,\qquad\qquad\qquad\quad\,\,\,4k_4(2+k_1)<k_3^2\label{ivham2}\\
&&\hspace{-1.1cm}\mbox{\bf(ivc)}\,\,H
=\displaystyle\log\left[\frac{w^{k_1}}{p}\right]+\frac{k_3pw^2}{2(2+k_1)},\qquad\qquad\qquad\qquad\qquad\qquad\quad\,\,\,4k_4(2+k_1)=k_3^2\label{ivham3}
\end{eqnarray}
where $p$ is the canonical momenta defined by
\begin{eqnarray}
&&\hspace{-1.1cm}\mbox{\bf(iva)}\,\,p=
\frac{2(2+k_1)}{w^2\sqrt{4(2+k_1)k_4-k_3^2}}\tan^{-1}\bigg[\frac{k_3w^2+2(2+k_1)w'}
{w^2\sqrt{4(2+k_1)k_4-k_3^2}}\bigg],\qquad 4k_4(2+k_1)>k_3^2\\
&&\hspace{-1.1cm}\mbox{\bf(ivb,c)}\,\,p=
\frac{w^{(2-r)k_1}}
{\left(w'+\frac{k_3(r-1)w^2}{r(2+k_1)}\right)^{r-1}},\qquad\qquad\qquad\qquad\qquad
\qquad\qquad\quad 4k_4(2+k_1)\le k_3^2
\end{eqnarray}

The existence of the time independent Hamiltonian confirms that the system  (\ref{eqiv}) is
an integrable one.  However, in the following we briefly point out the method of integrating the underlying Hamilton equations of motion
associated with the Hamiltonians (\ref{ivham1}) -
(\ref{ivham3})

\subsection{\bf General solution }
\label{caseivgensol}
To derive the general solution, we use the following
canonical transformations.

{\bf{Case {\bf (iv a)}: Parametric choice $k_3^2<4k_4(2+k_1)$}}

By introducing the canonical transformation, $w=\frac{U}{P}$ and $p=\frac{P^2}{2}$, where
$U$ and $P$ are new canonical variables,
we transform the Hamiltonian (\ref{ivham1}) (for the choice $k_3^2<4k_4(2+k_1)$) to the form
\begin{eqnarray}
H\equiv E=4(2+k_1)\log\left[\left(\frac{U}{P}\right)^{2+k_1}
\sec\left[\frac{U^2\sqrt{4(2+k_1)k_4-k_3^2}}{4(2+k_1)}\right]\right]-k_3U^2.
\label{tidcanh2}
\end{eqnarray}
The underlying canonical equations of motion then become
\begin{subequations}
\label{eq39}
\begin{eqnarray}
&&\hspace{-0.5cm}U'=\frac{-4(2+k_1)^2}{P},\qquad\qquad \left('=\frac{d}{dz}\right)\label{tidcaneq2}\\
&&\hspace{-0.5cm} P'=2k_3U-\frac{4(2+k_1)^2}{U}-2U\left(\sqrt{4(2+k_1)k_4-k_3^2}\right)
\tan\left[\frac{U^2\sqrt{4(2+k_1)k_4-k_3^2}}{4(2+k_1)}\right].
\end{eqnarray}
\end{subequations}

Eq. (\ref{eq39}) can be solved in the following way.
Expressing $P$ in terms of $\dot{U}$, by using the Eq. (\ref{tidcaneq2}), and substituting it in
 (\ref{tidcanh2}) we obtain
\begin{eqnarray}
E=4(2+k_1)\log\left[\mu_1(UU')^{2+k_1}\sec\left[\frac{U^2\sqrt{4(2+k_1)k_4-k_3^2}}
{4(2+k_1)}\right]\right]-k_3U^2,\label{eq41}
\end{eqnarray}
where $\mu_1=\frac{4^{-(2+k_1)}(-1)^{k_1}}{(2+k_1)^{2(2+k_1)}}$ and $E$ is an arbitrary constant.

By splitting $U'$ and $U$ in (\ref{eq41}),
\begin{eqnarray}
U'=\frac{\mbox{exp}\left[\frac{E+k_3U^2}{4(2+k_1)^2}\right]}
{U\left(\mu_1\sec\left[\frac{U^2\sqrt{4(2+k_1)k_4-k_3^2}}{4(2+k_1)}\right]\right)^{\frac{1}{(2+k_1)}}},
\end{eqnarray}
and integrating the above expression, we get
\begin{eqnarray}
z-z_0=&&F\left[\frac{1}{2+k_1},\frac{4a_1+ik_3}{8a_1(2+k_1)},
\frac{4a_1(5+2k_1)+ik_3}{8a_1(2+k_1)},-e^{2ia_1U^2}\right]\nonumber\\
&&\times\frac{(2+k_1)2^{\frac{3+k_1}{2+k_1}}e^{\frac{-(E+k_3U^2)+2ia_1U^2}{2(2+k_1)}}}{(k_3-4ia_1)
\mu_1^{\frac{1}{2+k_1}}}
\end{eqnarray}
where $a_1=\frac{\sqrt{4(2+k_1)k_4-k_3^2}}{4(2+k_1)}$ and $F$ is the hypergeometric function{\footnotesize$^{2}$}.\\
{\bf{Case {\bf (iv b)}: Parametric choice $k_3^2>4k_4(2+k_1)$}}

The Hamiltonian (\ref{ivham2}) (for the choice $k_3^2>4k_4(2+k_1)$) can be rewritten in terms of the new
canonical variables, $w=\frac{U}{P}$ and $p=\frac{P^2}{2}$, as
\begin{eqnarray}
H=2(2+k_1)rr_{12}\left(P^{2+k_1}U^{-k_1}\right)^{r_{12}}
-2^{r_{12}}k_3(r-1)U^2,\label{tidcanh1}
\end{eqnarray}
where we have defined $r_{12}=\frac{(r-1)}{(r-2)}$.
The Hamilton equations of motion corresponding to the above Hamiltonian are
\begin{subequations}
\label{eq44}
\begin{eqnarray}
&&U'=\frac{2(2+k_1)^2rr_{12}^2}{P}\left(\frac{P^{2+k_1}}
{U^{k_1}}\right)^{r_{12}},\label{tidcaneq1}\\
&&P'=2^{r_{12}}k_3(r-1)2U+\frac{2k_1(2+k_1)rr_{12}^2}{U}
\left(\frac{P^{(2+k_1)}}{U^{k_1}}\right)^{r_{12}}.
\end{eqnarray}
\end{subequations}
Now we integrate Eq. (\ref{tidcaneq1}) by
following the same analogy described in the previous subcase.  First
we rewrite Eq. (\ref{tidcaneq1}) for $P$ and obtain
\begin{eqnarray}
P=\left(\frac{U^{k_1r_{12}}U'}{2(2+k_1)^2rr_{12}^2}\right)^{\frac{1}{(2+k_1)r_{12}-1}}.\label{tidcanp1}
\end{eqnarray}
\begin{eqnarray}
H=\mu_2U^{m_1}U'^{m_2}+\mu_3U^2\equiv E,\label{eq48}
\end{eqnarray}
where
\begin{eqnarray}
&&\mu_2=\frac{2(2+k_1)rr_{12}}{\left(2(2+k_1)^2rr_{12}^2\right)^{m_3(2+k_1)r_{12}}}
,\,\,\,\,\mu_3=-2^{r_{12}}k_3(r-1),\,
\nonumber\\
&&m_1=k_1\left((2+k_1)m_3-r_{12}\right),\qquad \quad m_2=r_{12}(2+k_1)m_3.\nonumber
\end{eqnarray}

From (\ref{eq48}) one can express
\begin{eqnarray}
U'=\left(\frac{E-\mu_3U^2}{\mu_2U^{m_1}}\right)^{\frac{1}{m_2}}.
\label{eq47a}
\end{eqnarray}
Integrating (\ref{eq47a})  we get
\begin{eqnarray}
z-z_0=\frac{m_2U}{m_1+m_2}
\left(\frac{\mu_2U^{m_1}}{E}\right)^{\frac{1}{m_2}}F\left[\frac{m_1+m_2}{2m_2},\frac{1}{m_2},\frac{m_1+3m_2}{2m_2},
\frac{\mu_3U^2}{E}\right],
\end{eqnarray}
where $F$ is the hypergeometric function{\footnotesize$^{2}$} and $t_0$ is an integration constant.

{\bf{Case {\bf (iv c)}: Parametric choice $k_3^2=4k_4(2+k_1)$}}

We use the same canonical transformation $w=\frac{U}{P}$, $p=\frac{P^2}{2}$
and rewrite the Hamiltonian (\ref{ivham3}) with the parametric choice $k_3^2=4(2+k_1)$ as
\begin{eqnarray}
H=\log\left[\frac{U^{k_1}}{P^{k_1}}\right]-\log[P^2]
+\frac{k_3}{4(2+k_1)}U^2.\label{tidcanh3}
\end{eqnarray}
The associated canonical equations of motion now become
\begin{subequations}
\begin{eqnarray}
&&U'=-\frac{1}{P}(2+k_1),\label{tidcaneq3}\\
&&P'=-\left(\frac{k_1}{U}+\frac{2k_3U}{4(2+k_1)}\right).
\end{eqnarray}
\end{subequations}
Rewriting (\ref{tidcaneq3}) for $P=\frac{(2+k_1)}{-U'}$ and
substituting the latter into (\ref{tidcanh3}) we get
\begin{eqnarray}
H=\log[U^{k_1}(-4U')^{2+k_1}]+\frac{k_3U^2}{4(2+k_1)}\equiv E
\end{eqnarray}
which in turn can be brought to the form
\begin{eqnarray}
U'=-\frac{1}{4}
\left(U^{-k_1}\mbox{exp}[E-\frac{k_3}{4(2+k_1)}U^2]\right)^{\frac{1}{(2+k_1)}}.
\label{53a}
\end{eqnarray}
Now integrating the above equation (\ref{53a}) we get
\begin{eqnarray}
z-z_0=\frac{\tilde{E} U^{\frac{2(1+k_1)}{2+k_1}}}
{\left(\frac{-k_3U^2}{(2+k_1)^2}\right)
^{\frac{(1+k_1)}{2+k_1}}}
\Gamma\left[\frac{1+k_1}{2+k_1},\frac{-k_3U^2}{4(2+k_1)^2}\right],
\end{eqnarray}
where $\tilde{E}=-2^{\frac{4+4k_1}{2+k_1}}\mbox{exp}\left(\frac{-E}{2+k_1}\right)$ and
$\Gamma$ is the gamma function{\footnotesize$^{24}$}.

We summarize the results obtained in this section.  Solving the determining equations
(\ref{lin02})-(\ref{lin04}) we find that the system (\ref{eq6}) admits time dependent integrals for four parametric
choices and their respective equations are (\ref{caseia,b}),(\ref{td2-1}), (\ref{td3eq}) and (\ref{eqiv}).  For (\ref{caseia,b}) explicit solution is deduced by
integrating the corresponding time dependent integral of motion.  Using suitable
variable transformations, solutions of Eq. (\ref{td2-1}), Eq. (\ref{td3eq}) and Eq. (\ref{eqiv}) are
found in an implicit form.
\section{Connection with 2D Lotka-Volterra system (LV)}
\label{connection}
The detailed study made on the integrability of the second order
ODE (\ref{eq6}) in the previous sections helps one to identify the
dynamics of 2D-LV system which we find to be related to the second
order ODE (\ref{eq6}) under appropriate choice of $k_i$s.  The
2D-LV system
\begin{subequations}
\begin{eqnarray}
&&\dot {x}=x(a_{1}+b_{11}x+b_{12}y),\nonumber\\
&&\dot {y}=y(a_{2}+b_{21}x+b_{22}y),\qquad\qquad\qquad (\ref{eq1})\nonumber
\end{eqnarray}
\end{subequations}
models the the population dynamics of two interacting
species{\footnotesize$^{3}$}.  Interestingly Lotka-Volterra systems arises in
other branches of physics also such as the coupling of
waves in laser physics{\footnotesize$^{25}$} and the evolution of electrons, ions and
neutral species in plasma physics. In hydrodynamics they
model the convective instability in the Benard problem{\footnotesize$^{26}$}.
Similarly, they appear in the interaction of gases in a background
host medium{\footnotesize$^{27}$}. In the theory of partial differential equation
they can be obtained as a discretized form of the Korteweg-de
Vries equation{\footnotesize$^{28}$}.

Since the 2D-LV system is a planar dynamical system it is also
being thoroughly investigated from a mathematical point of view.
Due to the multi-faceted importance of the 2D-LV system, several
in-depth and independent studies have been made to classify the
integrable cases{\footnotesize$^{7-19}$}.
Integrals of motion of the 2D-LV system (\ref{eq1}) have been studied for several
parametric choices all of which reduces to any one of the following 3 parametric choices or subcases thereof :
\begin{eqnarray}
a_2&=&\frac{a_1b_{22}(b_{11}-b_{21})}{b_{11}(b_{12}-b_{22})},\\
b_{21}&=&\frac{b_{11}b_{22}}{b_{12}},\\
a_1&=&a_2.
\end{eqnarray}
for which integrals of motion have been explicitly deduced{\footnotesize$^{9}$}.
Interestingly, we find that all these three cases (see LV 1, LV 8, LV 15 below), in addition to
several subcases of the above parametric choices,
and their associated integrals of motion can be deduced from the
results of Eq. (\ref{eq6}) straightforwardly.

In the following we show that Eq. (\ref{eq1}) can be transformed to the form (\ref{eq6}) and thus the integrable cases 
of (\ref{eq6}) can be correlated with the integrable cases of
(\ref{eq1}).
\subsection{Transformation}
\label{sec12}
\n To transform the system (\ref{eq1}) to the form of Eq. (\ref{eq6}), first we
rewrite Eq. (\ref {eq1a}) for the variable $y$ as
\begin{eqnarray}
y=\frac{1}{b_{12}}\left(\frac{\dot {x}}{x}-b_{11}x-a_{1}\right),\,\,b_{12}\ne0\label {eq5a}
\end{eqnarray}
Then we substitute the latter into Eq. (\ref {eq1b}) and obtain the following equation,
\begin{eqnarray}
&&\hspace{-1cm}\ddot{x}-\left(1+\frac{b_{22}}{b_{12}}\right)\frac{\dot{x}^{2}}{x}+
\left((2b_{11}\frac{b_{22}}{b_{12}}-b_{11}-b_{21})x+(2a_{1}\frac{b_{22}}
{b_{12}}-a_2)\right)\dot{x}+\left(b_{21}b_{11}-\frac{b_{22}}{b_{12}}b_{11}^{2}\right)x^{3}
\nonumber\\&&\quad
+\left(b_{11}a_2+b_{21}a_{1}-2a_{1}b_{11}\frac{b_{22}}{b_{12}}\right)x^{2}
+\left(a_1a_2-\frac{b_{22}}{b_{12}}a_{1}^{2}\right)x=0.
\label {eq5}
\end{eqnarray}
which is of the same form as (\ref{eq6}).
Now comparing (\ref{eq5})  with Eq. (\ref{eq6}) we find the parameters are connected in the following way
\begin{eqnarray}
 &&\hspace{-0.5cm}k_1=-(1+\frac{b_{22}}{b_{12}}),
\quad
k_{2}=(2a_{1}\frac{b_{22}}{b_{12}}-a_2),
\quad
k_{3}=2b_{11}\frac{b_{22}}{b_{12}}-b_{11}-b_{21},
\nonumber\\
 &&\hspace{-0.5cm}k_{4}=(b_{21}b_{11}-\frac{b_{22}}{b_{12}}b_{11}^{2}),\quad
k_{5}=(b_{11}a_2+b_{21}a_{1}-2a_{1}b_{11}\frac{b_{22}}{b_{12}}),
\quad k_{6}=(a_1a_2-\frac{b_{22}}{b_{12}}a_{1}^{2}).\label {eq7}
\end{eqnarray}
One may note that both the Eqs. (\ref{eq6}) and (\ref{eq5}) contain six parameters and thus one ends up with six
relations connecting them.  Here we emphasize that one can obtain the results of the LV equation from the results of Eq. (\ref{eq6}) straightforwardly, while
it is difficult to deduce all the results pertaining to Eq. (\ref{eq6}) from the known results of LV equation.  This is illustrated with an example.
One of the integrable parametric choices of the LV equation (\ref{eq1}) is
$a_2=\frac{a_1b_{22}(b_{11}-b_{21})}{b_{11}(b_{12}-b_{22})}$.  Upon substituting the above
parametric choice in the relations (\ref{eq7}) and solving for $k_i$'s we get,
$k_5=\frac{k_2k_4(3+2k_1)}{k_3(1+k_1)}$, $k_6=\frac{k_4k_2^2(2+k_1)}{k_3^2(1+k_1)}$ and
$b_{11}=\frac{-k_3\pm\sqrt{k_3^2-4(2+k_1)k_4}}{2(2+k_1)}$ where $b_{11}$ is a real and arbitrary constant.  We find that
this $b_{11}$ is equivalent to $r=\frac{-k_3\pm\sqrt{k_3^2-4(2+k_1)k_4}}{2(2+k_1)}$ of the parametric choice $k_3^2>4(2+k_1)$.
However, we have obtained integrals of motion
for all the three parametric choices $k_3^2\ge4(2+k_1)$ and $k_3^2<4(2+k_1)$ which cannot be obtained going back from the results of LV.
\section{Results in terms of Lotka-Volterra equation parameters}
\label{seclv}
\label{resultcon}
Now we rewrite the results obtained in Secs. \ref{sectid} - \ref{sectd} in terms
of the LV parameters using the relation (\ref{eq7}).  In total we obtain 16 parametric choices in terms of the LV parameters 
(designated as LV 1 - LV 16) and the results are
tabulated in Tables \ref{table3} and \ref{table4}, corresponding to time independent and time dependent integrals, respectively.  
It may be noted that out of the 16 parametric choices, 15 of them reduce to any one of the
three parametric choices LV 1, LV 8 and LV 15 (given in the last columns of Tables \ref{table3} and \ref{table4}), while the remaining one (LV 6) is the uncoupled case.
  Also we note that no new parametric choice is obtained other than the ones already reported in the literature {\footnotesize$^{9,12,13,15,16}$} as far as the Lotka-Volterra
system (\ref{eq1}) is concerned.
\vskip 6pt
\begin{sidewaystable}
\caption{Parametric cases of Eq. (\ref{eq6}) possessing time independent integrals of motion in terms of the LV
parameters}
\noindent\begin{tabular}{llll} \hline
 Case\,\, & Parametric restrictions
\quad& \qquad Parametric restrictions in terms of LV parameters& \qquad Integrable choices of the LV equation
 \\
&\qquad in Eq. (\ref{eq6})& &\\
\hline\\
{\bf (i)}&$k_5=\frac{k_2k_4(3+2k_1)}{k_3(1+k_1)}$& \qquad
(i) $(a_2b_{11}(b_{12}-b_{22})+a_1(b_{21}-b_{11})b_{22})
=0,$ & {\bf LV 1} $
a_2=\frac{a_1b_{22}(b_{11}-b_{21})}{b_{11}(b_{12}-b_{22})}$\\
            &$k_6=\frac{k_4k_2^2(2+k_1)}{k_3^2(1+k_1)}$.&\qquad (ii) $a_1=a_2$, $b_{21}=0$ & {\bf LV 2} $a_1=a_2,\,b_{21}=0$\\
&&\\
{\bf (ii)} &$k_2=0,\,k_3=0$& \qquad $a_2-\frac {2a_1b_{22}}{b_{12}}=0$,\,
$ b_{11}+b_{21}-\frac {2b_{11}b_{22}}{b_{12}}=0$ &{\bf LV 3} $a_2=\frac{2a_1b_{22}}
{b_{12}},\, b_{21}=b_{11}\left (\frac{2b_{22}}{b_{12}}-1\right )$\\
&&\\
{\bf (iii)} &$k_1=0,\,k_2=0,$ &\qquad  $\frac{b_{22}}{b_{12}}+1=0,\,2a_1\frac{b_{22}}{b_{12}}-a_2=0,$ &{\bf LV 4} $b_{12}=-b_{22},\,a_1=-\frac{a_2}{2},\,b_{21}=0$
\\
 & $k_5=0$& \qquad $b_{11}a_2+b_{21}a_1-2a_1b_{11}\frac{b_{22}}{b_{12}}=0$ &{\bf LV 5} $b_{12}=-b_{22},\,a_1=a_2=0$\\
 & &\\
\hline
\end{tabular}
\label{table3}
\end{sidewaystable}

\begin{sidewaystable}
\caption{Parametric cases of Eq. (\ref{eq6}) possessing time dependent integrals of motion in terms of the LV
parameters}
\noindent\begin{tabular}{llll}
\hline Case\,\, & Parametric restrictions &\qquad Parametric restrictions in terms of LV parameters & Integrable choices of the LV equation
\\
&\qquad in Eq. (\ref{eq6})&&\\
\hline
 (i)&$k_4=\frac{(1+k_1) k_3^2}{(3+2 k_1)^2}$, & \qquad$\frac{b_{21}(b_{11}(b_{12}-2b_{22})+b_{21}b_{22})}{b_{12}-2b_{22}}=0,$& {\bf LV 6} $b_{21}=0,\,\,b_{12}\ne 0$\\
     &$k_5=\frac{k_2k_3}{3+2k_1}$ &  \qquad$\frac{b_{21}(a_1-a_2)}{b_{12}-2b_{22}}=0$ & {\bf LV 7} $a_1=a_2,\,b_{11}=-\frac{b_{21}b_{22}}{b_{12}-2b_{22}}$\\
&&&\\
(ii)& $k_4=0,$& \qquad$b_{11}b_{21}-\frac{b_{11}^2b_{22}}{b_{12}}=0,$ & {\bf LV 8} $b_{21}=\frac{b_{11}b_{22}}{b_{12}}$\\
&$k_5=\frac{k_3(k_2\pm\omega)}{2 (2+k_1)}$& \qquad$\frac{a_2b_{12}(b_{11}-b_{21})+2a_1(b_{12}b_{21}-b_{11}b_{22})\pm
a_2(b_{11}b_{12}+b_{12}b_{21}-2b_{11}b_{22})}{b_{12}-b_{22}}=0$ & {\bf LV 9} $a_1=a_2,\,b_{11}=0$,\,\,\mbox{\bf LV 10} $a_1=0,\,b_{11}=0$\\
&&&\\
(iii)& $k_3=0,\,k_4=0,$& \qquad$\frac{2 b_{11}b_{22}}{b_{12}}-b_{21}-b_{11}=0,\,b_{11}b_{21}-\frac{b_{11}^2 b_{22}}{b_{12}}=0,$& {\bf LV 11}  $a_1=2a_2,\,b_{11}=0,\,b_{21}=0$\\
& $k_6=\frac{2(3+2 k_1)k_2^2}{(5+4 k_1)^2}$& \qquad$\frac{(a_1-2a_2)(a_1b_{22}+2a_2b_{22}-a_2b_{12})}{b_{12}-4b_{22}}
=0$& {\bf LV 12} $a_1=\frac{a_2(b_{12}-2b_{22})}{b_{22}},\,b_{11}=0,\,b_{21}=0$,\\
&&&\mbox{\bf LV 13} $a_1=2a_2,\,b_{11}=b_{21},\,b_{12}=b_{22}$\\
&&&\mbox{\bf LV 14} $a_1=-a_2,\,b_{11}=b_{21},\,b_{12}=b_{22}$\\
&&&\\
(iv)&$k_5=\frac{k_2k_3}{3+2k_1}$,& \qquad$\frac{b_{21}(a_1-a_2)}{b_{12}-2b_{22}}=0$,& {\bf LV 15} $a_1=a_2\label{LV15}$\\
    &$k_6=\frac{(2+k_1)k_2^2}{(3+2k_1)^2}$ & \qquad$\frac{(a_1-a_2)
(a_1b_{22}+a_2(b_{22}-b_{12}))}{b_{12}-2b_{22}}=0$& {\bf LV 16} $a_1=\frac{a_2(b_{12}-b_{22})}{b_{22}},\,b_{21}=0$\\
& & &\\
\hline
\end{tabular}
\label{table4}
\end{sidewaystable}

To deduce the integrable choices LV 1 - LV 16 from the results of Eq. (\ref{eq6}) one can derive the corresponding
integrals of motion for each one of the cases from the results of the second order equation and the relation (\ref{eq7}).  As we pointed out earlier, the integrability
of the parametric choices obtained in this analysis have already been established in Ref. [9,12,13,15,16].
In the following we illustrate the procedure to
deduce
the time independent integral from the results of the second order equation (\ref{eq6}) for the parametric choices LV 1 and LV 2.  Integrals of motion for the remaining cases can be
deduced similarly and the procedure is straightforward.  Therefore we do not present the details here.  

Considering the integrals of motion of the second order system (\ref{eq6})
we note that case (i) have three types of
integrals (vide Eqs. (\ref{integral6})) depending on the values of the parameters.  To
rewrite these integrals for the first order LV system (\ref{eq1}) first we
check whether the LV parametric relations LV 1 and LV 2 are consistent with
these conditions. While verifying this we find that both the LV
systems are subcases of the overdamped parametric choices $k_3^2>4k_4(2+k_1)$, see Eq. (\ref{integral6}).

Once the respective integral has been identified
 then one
can replace the variable $\dot{x}$ in terms $x$ and $y$ (vide eq (\ref{eq1}))
in (\ref{integral6}).  The integrals of motion
for the above two LV cases turn out to be, respectively,
\begin{eqnarray}
&&\hspace{-0.7cm}\mbox{\bf LV 1}\,\,I_1=yx^{\frac{b_{22}(b_{21}-b_{11})}
{b_{11}(b_{12}-b_{22})}}(a_1(b_{11}-b_{21})+b_{11}(b_{11}-b_{21})x
+b_{11}(b_{12}-b_{22})
y)^{\frac{b_{22}b_{11}-b_{12}b_{21}}{b_{11}(b_{12}-b_{22})}}\\
&&\hspace{-0.7cm}\mbox{\bf LV 2}\,\,I_1=(a_1+b_{22}y)(b_{11}x+
(b_{12}-b_{22})y)^{\frac{b_{22}}{b_{12}-b_{22}}}x^{\frac{-b_{22}}{b_{12}-b_{22}}}
\end{eqnarray}
We also note that the well known LV equation
\begin{subequations}
\begin{eqnarray}
&&\dot{x}=x(a_1-b_{21}y),\\
&&\dot{y}=y(a_2+b_{21}x),
\end{eqnarray}
\label{classical}
\end{subequations}
is a subcase of LV 1.  The integral of motion corresponding to (\ref{classical}) is deduced from the integral (\ref{integral6}) (with $k_1=-1$) as
\begin{eqnarray}
I_1=b_{21}(x+y)+a_2\log[x]-a_1\log[y]).
\end{eqnarray}
For the remaining cases LV 3- LV 16, similar analysis can be performed straightforwardly.


\section{Conclusion}
\label{seccon}
In this paper, we have investigated the integrability properties of Eq. (\ref{eq6}) and shown
that it admits a set of integrable parametric choices.  To identify them we have divided
our analysis into two categories, that is systems which admit time independent first integrals and equations which
possess time dependent integrals.  After carrying out the detailed analysis we found that there exists a new equation
which admits time independent integral.  To interpret this integral as a Hamiltonian we first
deduce the corresponding Lagrangian and then construct the Hamiltonian using the Legendre transformation.  Since
we have identified a conservative Hamiltonian description for a dissipative system, we expect the study can be
extended to the quantum case as well in future.  The other two systems which admit time independent integrals are
already known in the literature.  However we have also given the Hamiltonian description for both of them.

We then moved on to identify the systems which admit time dependent integrals.  Our results show that there exist
four integrable cases in (\ref{eq6}) that admit time dependent integrals.  We have also reported the explicit forms of
these integrals.  For the first three equations we have also found the general solution from these integrals. Since the integral
of the
fourth equation turned out to be a very complicated one it became difficult to integrate it straightforwardly.  So first
we have transformed the
time dependent integral into a time independent one.  Then from the latter we identified a Hamiltonian.  We then
introduced a canonical transformation to this Hamiltonian and transformed the latter into a relatively simpler Hamiltonian.  This Hamiltonian
has been integrated to obtain the general solution.

We have transformed the
identified integrable choices of the second order equation to the LV system.  Out of the 16 LV parametric choices obtained,
15 reduces to any one of the three parametric choice LV 1, LV 8, LV 15.  The 16$^{th}$ one, namely LV 6, is an uncoupled case.
Interestingly our results reproduce
all the known integrable cases of the LV system in the literature.

\section{Acknowledgements}
The work forms a part of a research project of MS and an IRPHA
project of ML sponsored by the Department of Science \& Technology
(DST), Government of India.  ML is also supported by a DST Ramanna
Fellowship.
\vskip 14pt
\begin{tabular}{p{.15cm}p{14cm}}
{\footnotesize$^{1}$}  &
E. Kamke: \emph{Differentialgleichungen Losungsmethoden und Losungen},
Stuggart: Teubner, 1983.\\
{\footnotesize$^{2}$}  &
George M. Murphy \emph{Ordinary Differential Equations and Their Solutions},
An East-West Editon 1969, New Delhi.\\
{\footnotesize$^{3}$}  &
J.D Murray \emph{Mathematical Biology} (Springer-Verlag, New York, 1989)\\
{\footnotesize$^{4}$}  &
M. Lakshmanan and S. Rajasekar \emph{Nonlinear Dynamics: Integrability
Chaos and Patterns} (Springer-Verlag, New York, 2003)
\end{tabular}
\newpage
\begin{tabular}{p{.15cm}p{14cm}}
{\footnotesize$^{5}$}  &
V. K.Chandrasekar, M. Senthilvelan and M. Lakshmanan,
Proc. R. Soc. London A {\bf 461}, 2451 (2005)\\

{\footnotesize$^{6}$}  &
V. K.Chandrasekar, M. Senthilvelan and M. Lakshmanan, J. Nonlinear
Math. Phys. {\bf 12}, 184 (2005)\\

{\footnotesize$^{7}$}  &
D.D Hua, L. Cairo, M.R. Feix, K.S. Govinder and P.G.L. Leach,
Proc. R. Soc. London A {\bf 452}, 859 (1996)\\
{\footnotesize$^{8}$}  &
P.L. Sachdev and Sharadha Ramanan, J. Math. Phys. {\bf 34}
4025 (1992)\\
{\footnotesize$^{9}$}  &
L. Cairo,M. R. Feix and J. Goedert, Phys. Lett. A {\bf 140}, 421 (1989)\\

{\footnotesize$^{10}$}  &
L. Cairo, and M.R. Feix, J. Math. Phys. {\bf 33} 2440 (1992)\\
{\footnotesize$^{11}$}  &
M.A. Almeida, M.E. Magalhaes and I.C Moreira, J. Math. Phys. {\bf 36} 1854 (1995)\\
{\footnotesize$^{12}$}  &
L. Cairo, J. Llibre,  J. Phys. A {\bf 33}, 2407 (2000)\\
{\footnotesize$^{13}$}  &
L. Cairo, M.R. Feix and J. Llibre J. Math. Phys. {\bf 40} 2074 (1999)\\
{\footnotesize$^{14}$}  &
L. Cairo and M.R. Feix J. Phys. A {\bf 25}, L1287 (1992)\\
{\footnotesize$^{15}$}  &
J. Llibre and C. Valls, J. Math. Phys. {\bf 48}, 033507 (2007)\\
{\footnotesize$^{16}$}  &
L. Cairo, H. Giacomini and J. Llibre,  Rend. Circ. Mat. Palermo {\bf 52}, 389 (2003)\\
{\footnotesize$^{17}$}  &
J. Moulin Ollagnier, Bull. Sci. Math. {\bf 121}, 463 (1997)\\
{\footnotesize$^{18}$}  &
J. Moulin Ollagnier, Bull. Sci. Math. {\bf 123}, 437 (1999)\\
{\footnotesize$^{19}$} 
& J. Moulin Ollagnier, Qualitative
Theory of Dynamical Systems, {\bf 2} 307 (2001)\\
{\footnotesize$^{20}$}  &
V. K. Chandrasekar, S.N. Pandey, M. Senthilvelan and M. Lakshmanan,
 J. Math. Gen. {\bf 47} 023508
(2006)\\
{\footnotesize$^{21}$}  &
V K Chandrasekar, M Senthilvelan and M Lakshmanan Phys. Rev. E {\bf 72},
066203 (2005)\\
{\footnotesize$^{22}$}  &
R Gladwin Pradeep, V K Chandrasekar, M Senthilvelan and M Lakshmanan J. Math. Phys. {\bf 50}, 052901 (2009)\\
{\footnotesize$^{23}$}  &
R. Iacono, J. Phys. A : Math. Theor. {\bf 41}, 068001, (2008)\\
{\footnotesize$^{24}$}  &
I.S Gradshteyn and I.M. Ryzhik, \emph {Table of Integrals, Series and Products}
(Academic Press, London, 1980)\\
{\footnotesize$^{25}$}  &
W. E. Lamb, Phys. Rev. A 134, 1429 (1964)\\
{\footnotesize$^{26}$}  &
F.H. Busse, \emph{Synergetics} (Springer, Berlin, 1978)\\
\footnotesize$^{27}$  &
R. Lupin and G. Spiga, Phys. Fluids 31, 2048 (1988)\\
\footnotesize$^{28}$  &
O.I. Bogoyavlensky, Phys. Lett. A 134, 34 (1988)
\end{tabular}
\end{document}